\documentclass{aa}
\usepackage{graphicx}
\begin{document}
   \title{Gamma-rays from the pulsar wind nebulae}

   \author{W. Bednarek \& M. Bartosik}

   \offprints{bednar@fizwe4.fic.uni.lodz.pl}

   \institute{Department of Experimental Physics, University of \L \'od\'z,
        ul. Pomorska 149/153, 90-236 \L \'od\'z, Poland
             }

   \date{Received ; accepted}

   \abstract{
We investigate the radiation processes inside supernova remnants which are powered
by young pulsars. 
Using recent model for particle acceleration by the pulsar wind nebulae (PWNe),
in which positrons gain energy in the process of resonant scattering by the heavy nuclei, 
we construct the time dependent radiation model for the PWNe. In this model,
the spectra of relativistic particles, injected inside the nebula, depend on time
due to the evolution of the pulsar parameters.
Applying a simple model for the evolution of the PWNa,
the equilibrium spectra of leptons and nuclei inside the nebula are determined 
as a function of time, taking into account the energy losses of particles on different 
processes and their escape from the nebula. We calculate the multiwavelength photon 
spectra produced by leptons and nuclei and compare them with the observations of the 
PWNe for which the TeV $\gamma$-ray emission has been reported, i.e. the Crab Nebula, 
the Vela Supernova Remnant, and the nebula around PSR 1706-44. It is found that the 
emission from the Crab Nebula can be well fitted by the composition of the $\gamma$-ray 
emission produced by leptons (below $\sim 10$ TeV) and nuclei ($\sim 10$ TeV). 
The model is further tested by successful fitting of the high energy spectrum from the 
Vela SNR. In this case, the observed $\gamma$-ray emission is mainly due to leptons and 
the contribution of $\gamma$-rays from decay of neutral pions, produced in collision of 
nuclei, is significantly lower.
However, considered model does not give good fitting to the emission from PSR1706-44, 
for the likely parameters of this source, unless additional target for 
relativistic leptons is present inside the nebula, e.g. the thermal infrared emission.
Based on the knowledge obtained from these fittings, we
predict the $\gamma$-ray fluxes in the TeV energy range from other PWNe, which are 
promissing TeV $\gamma$-ray sources due to their similarities to the $\gamma$-ray 
nebulae, i.e. MSH15-52 (PSR 1509-58), 3C58 (PSR J0205+6449), and CTB80 (PSR 1951+32).
Possible detection of these sources by the new generation of Cherenkov telescopes
is discussed.   

   }

   \maketitle
%
%________________________________________________________________

%
%
\section{Introduction}

Young pulsars, born in the supernova explosions, create relativistic winds which 
in the early stage of evolution interact with the supernova ejecta creating the 
pulsar wind nebulae. These nebulae are filled with energetic particles which can
radiate in all range of electromagnetic spectrum. Up to now, several objects of this
type are observed (e.g. Chevalier~2003). Between them three nebulae, the Crab Nebula
and the nebulae around the Vela and PSR1706-44 pulsars, have been detected
in the TeV $\gamma$-rays (see for review e.g. Fegan~2001). The best studied object, 
the Crab Nebula, shows $\gamma$-ray emission up to 50 TeV with very well established 
spectrum, determined by several independent telescopes starting from 
the original detection by the Whipple group (Weekes et al.~1989).
Recently, another PWNa, MSH15-52 around the pulsar PSR 1509-58, has been marginally 
detected by the CANGAROO telescope (Sako et al.~2000).

It is widely argued that lower energy radiation in the PWNa is produced by leptons
in the magnetic field (the synchrotron radiation) and the higher energy part of the 
spectrum is produced by leptons in the inverse Compton scattering
of the low energy synchrotron, Microwave Background Radiation (MBR), or infrared photons.
Leptons are accelerated as a result of interaction of the pulsar wind with the nebula
(e.g. Kennel \& Coroniti~1984).  
First discussions of these radiation processes, with the application to the Crab Nebula, 
were given by Gould (1965) and Rieke \& Weekes (1969) and a more 
physical models were analyzed by Grindlay \& Hoffman (1971) and Stepanian (1980).
More recently, detailed modeling of the observed high energy emission from the Crab 
Nebula have been presented by e.g., de Jager \& Harding 1992, Atoyan \& Aharonian 1996, 
de Jager et al. 1996, and Hillas et al. 1998.
Also high energy processes in other PWNe have been discussed in more detail, 
concentrating on the sources from which detection of $\gamma$-ray photons 
have been claimed, e.g. nebulae around: PSR 1706-44 
(Aharonian, Atoyan \& Kifune~1997), PSR 1509-58 (Du Plessis et al.~1995), Vela pulsar
(De Jager et al.~1996b). For example, Aharonian et al.~(1997) consider in detail the case
of nebula around PSR 1706-44, assuming that leptons are injected continuously during the 
lifetime of the pulsar with the constant rate and the power law spectrum.

It is likely that also production of $\gamma$-rays in the interactions of hadrons with 
the matter of the supernova can contribute to the observed spectrum in its higher energy 
end, especially in the case of younger nebulae (see e.g. Cheng et al.~1990, Aharonian \& 
Atoyan~1996, and Bednarek \& Protheore~1997). 
In fact, the presence of heavy nuclei inside PWNe is quite likely since the significant 
part of the energy lost by the pulsar can be taken by relativistic iron nuclei.
These nuclei excite waves, just 
after the pulsar wind shock, which energy can be resonantly transfered to positrons 
(Hoshino et al.~1992). Apart from providing efficient mechanism for acceleration of 
leptons inside the PWNe, such model can also successfully explain the appearance 
of thin synchrotron filaments in the Crab Nebula (Gallant \& Arons~1994).

Based on the recent works by Arons and collaborators (see e.g. Arons~1998), 
we constrain the time dependent model for the acceleration, propagation, and
radiation of particles inside the PWNe. It is assumed that pulsar injects into the nebula 
relativistic heavy nuclei, with the Lorentz factors determined by the pulsar parameters, 
which evolve in time due to the pulsar rotational energy losses (see e.g. Gallant \& 
Arons~1994). Therefore, the injection spectra of leptons and nuclei 
are time dependent as well.
Leptons and nuclei are injected into the surrounding nebula, which parameters also evolve 
in time, due to expansion of the nebula. The parameters of the nebula are in turn 
determined by the 
initial kinetic energy of the supernova, the energy supply by the pulsar, and the 
surrounding medium. We calculate the equilibrium spectra of leptons and nuclei
inside the nebula as a function of time, taking into account radiation and adiabatic 
energy losses and escape of particles from the nebula. Then, we calculate
the photon spectra produced by these particles in processes, which are likely to 
contribute to the multiwavelength emission from the PWNe.

The paper is organized as follows. We constrain the time dependent model for the 
expending nebula in section 2, following the early ideas developed by Ostriker \& 
Gunn~(1971) and Rees \& Gunn~(1974). In section 3 the model for the acceleration of heavy 
ions and leptons inside the nebula is described. In Sect. 4 and 5, the equilibrium 
spectra of particles inside nebula are obtained and the radiation mechanisms are discussed. 
Finally, in Sect.~6, we confront such a model with the observations of 
the Crab Nebula, the Vela nebula, and the nebula around PSR 1706-44. From the comparison 
of calculated spectra with the observations of these nebulae, we derive 
some free parameters of the considered model in order to predict the level of 
$\gamma$-ray emission for other nebulae: 3C58 around PSR J0205+6449, 
CTB80 around PSR 1951+32, and MSH 15-52 around PSR 1509-58.

\section{Interaction of the pulsar with the supernova envelope}

The rotational energy of pulsars, formated during supernova explosion, can be
comparable (or even higher) to the kinetic energy of expending envelopes.
This energy is lost by the pulsar in the form of relativistic particles and  
electromagnetic waves, whose energy is absorbed by the envelope. Therefore, in 
the case of very young pulsars, the parameters of expending 
envelopes are determined by not only the initial parameters of the envelope and the 
surrounding medium but also by the energy loss rate of the pulsars.

Since in this paper we intend to consider the case of supernova exploding  
with arbitrary initial parameters for the neutron star and the 
envelope, the transfer of energy from the pulsar to the envelope has to 
be taken into account. In fact, the evolution of the supernova envelope under the 
influence of the pulsar has been considered soon after the pulsar discovery by 
Ostriker \& Gunn~(1971), (see also later works
by e.g. Pacini \& Salvati~(1973) or Reynolds \& Chevalier~(1984)). 
In order to determine the parameters of the envelope surrounding the pulsar, we follow 
general prescription presented in this first paper. 
Let us assume that at the moment of explosion the 
expansion velocity of the supernova envelope at its inner radius is $V_{\rm 0,Neb}$ and 
its initial mass is $M_{\rm 0,Neb}$. The expansion velocity, $v_{\rm Neb}(t)$, increases 
at the early stage, due to the additional supply of energy to the nebula by the pulsar, 
and decreases at the later stage, due to the accumulation of the surrounding matter. 
At an arbitrary time, t, it can be determined by analyzing the energy budget of the 
envelope by applying the simple equation, 
\begin{eqnarray}
{{M_{\rm Neb}(t)V_{\rm Neb}^2(t)}\over{2}} = {{M_{\rm 0,Neb}V_{\rm 0,Neb}^2}\over{2}}
+ \int^t_0L_{\rm pul}(t')dt',
\label{eq1}
\end{eqnarray}
\noindent
where 
\begin{eqnarray}
L_{\rm pul}(t) = B_{\rm s}^2 R_{\rm s}^6 \Omega^4/6c^3\approx 
3\times 10^{43}B_{12}^2P_{\rm ms}^{-4}~~{\rm erg~s}^{-1}, 
\label{eq2}
\end{eqnarray}
\noindent
is the rate of pulsar energy lost on emission of the dipole electromagnetic radiation,
$\Omega = 2\pi/P$, and the period of the pulsar $P = 10^{-3}P_{\rm ms}$ s
changes with time according to 
\begin{eqnarray}
P^2_{\rm ms}(t) = P_{\rm 0,ms}^2 + 2\times 10^{-9}tB_{12}^2, 
\label{eq3}
\end{eqnarray}
\noindent
where $P_{\rm 0,ms}$ is the initial period of the pulsar and $B =10^{12}B_{12}$
G is the strength of its surface magnetic field. 
The expending nebula increases the mass from the surrounding medium according to
\begin{eqnarray}
M_{\rm Neb}(t) = M_{\rm 0,Neb} + {{4}\over{3}}\pi \rho_{\rm sur} R_{\rm Neb}^3(t),
\label{eq4}
\end{eqnarray}
\noindent
where $\rho_{\rm sur}$ is the density of surrounding medium, and $R_{\rm Neb}(t)$ 
is the radius of expending envelope at the time, t, which depends on the 
expansion history of the nebula,
\begin{eqnarray}
R_{\rm Neb}(t) = \int_0^t V_{\rm Neb}(t')dt'.
\label{eq5}
\end{eqnarray}
$V_{\rm Neb}(t)$, and the average density of matter inside the nebula, 
$\rho_{\rm Neb}(t) = 3M_{\rm Neb}(t)/4\pi R_{\rm Neb}^3(t)$,
have been found numerically by solving the above set of Eqs. (1-5). These parameters 
are shown in Fig.~1 for different initial periods of the pulsar.

\begin{figure}
  \vspace{12.truecm}
\includegraphics{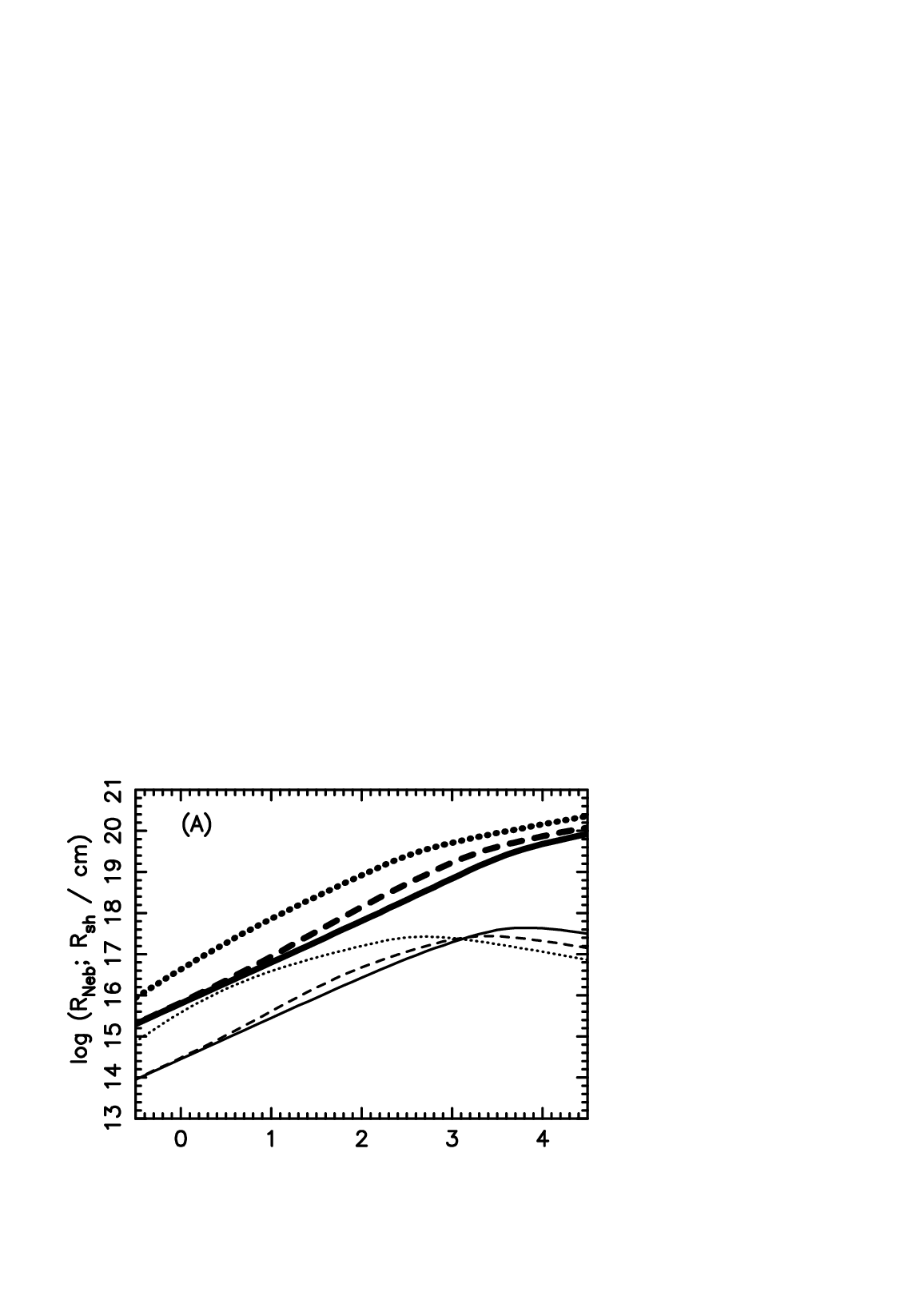}
\includegraphics{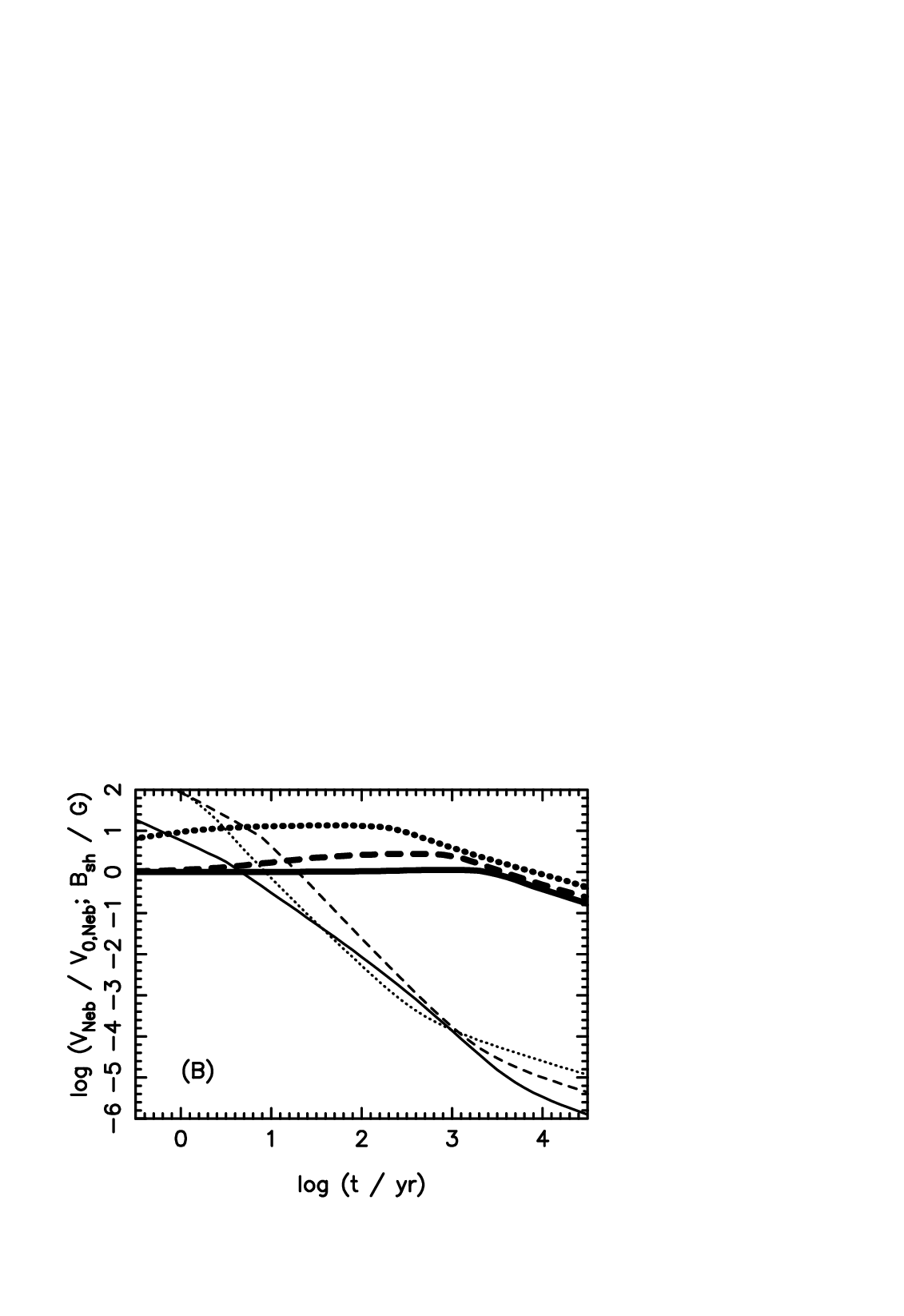}
  \caption{(A) The characteristic distances in the nebula are shown as a function of 
time: the radius of the nebula $R_{\rm Neb}$ (thick curves), 
the radius of the pulsar wind shock $R_{\rm sh}$ (thin curves), for the pulsar surface 
magnetic field $5\times 10^{12}$ G and its three initial periods: 1 ms (dotted curves), 
5 ms (dashed curves) and 20 ms (full curves). The 
initial expansion velocity of the nebula is 2000 km s$^{-1}$,  
its mass 4 M$_{\odot}$, and the density of surrounding medium is 0.1 cm$^{-3}$. 
(B) The expansion velocity of the nebula (thick curves) and the strength of 
the magnetic field at the shock region (thin curves) are shown as a function of time for 
the same initial parameters of the pulsar and nebula as in Figure (A).} 
\label{fig1}
\end{figure}

The pulsar loses energy in the form of relativistic wind which extends
up to the distance, $R_{\rm sh}$, at which the pressure of the wind is balanced 
by the pressure of expending nebula. We obtain the location of $R_{\rm sh}$ 
as a function of time by comparing the wind energy flux with the pressure of the outer 
nebula, which is in turn determined by the supply of magnetic energy to 
the nebula by the pulsar over its all lifetime (Rees \& Gunn~1974),
\begin{eqnarray}
{{L_{\rm pul}(t)}\over{4\pi R_{\rm sh}^2c}}\approx 
{{\int_0^t\eta L_{\rm pul}(t')dt'}\over{{{4}\over{3}}\pi R_{\rm Neb}^3}},
\label{eq6}
\end{eqnarray}
\noindent
where $\eta$ is the ratio of the magnetic energy flux to the total energy 
flux lost by the pulsar, measured at the location of the pulsar wind shock at 
$R_{\rm sh}$. The value of $\eta$ changes with the age of the pulsar. 
It is related to so called 'magnetization parameter', $\sigma$, which is the ratio 
of the magnetic energy flux to the particle energy flux lost by the pulsar by 
$\eta = \sigma/(1 + \sigma)$. The evolution of $\sigma$ with the parameters of the pulsar 
is found by interpolating between the values estimated for the Crab pulsar, $\sim 0.003$, 
and for the Vela pulsar, $\sim 1$ (see for details Eq. 16 and below in Bednarek \& 
Protheroe~2002). 

Knowing how the magnetic field depends on the distance from the pulsar, in the pulsar
inner magnetosphere and in the pulsar wind zone, we can estimate the strength of the 
magnetic field after the wind shock from, 
\begin{eqnarray}
B_{\rm sh}\approx 3\sigma B_{\rm pul}\left({{R_{\rm pul}}\over{R_{\rm lc}}}\right)^3
{{R_{\rm lc}}\over{R_{\rm sh}}},
\label{eq10}
\end{eqnarray}
\noindent
where $R_{\rm pul}$ and $B_{\rm pul}$ are the radius and the surface magnetic field 
of the pulsar. 

In our further considerations it is assumed that the typical parameters of the nebula
at the moment of supernova explossion are: the expansion velocity of of the bulk matter 
of the supernova $V_{\rm 0,Neb} = 2\times 10^3$ km s$^{-1}$,  
and its initial mass $M_{\rm 0,SN} = 4M_{\odot}$ (with good agreement with the 
parameters derived for the Crab Nebula, see Davidson \& Fesen~1985). 

\section{Acceleration of particles by the pulsar}

In this section we define the spectrum of particles which are injected into the nebula.
It is assumed that significant part of rotational energy of the pulsar is curried out by
iron nuclei, which are extracted from the surface of the neutron star. In fact, the 
cohesive 
and binding energy of the iron nuclei on the neutron star surface is not well known.
If it is of the order of 2-3 keV (for the magnetic field of a few $10^{12}$ G, see
Abrahams \& Shapiro~1991), then the iron nuclei can be thermally emitted for the surface
temperature $T\cong 3.5\times 10^5 B_{12}^{0.73}$ K (Usov \& Melrose~1995).
According to the standard cooling model of the neutron stars (Nomoto \& Tsuruta~1987),
their surface temperatures are $> 10^6$ K for the age $< 10^5$ yrs. Therefore, it is
possible that the neutron stars in the PWNe inject the iron nuclei.
These nuclei can be accelerated in the inner magnetosphere and/or the pulsar wind zone.
From normalization to the observations of the Crab pulsar,  
Arons and collaborators (see Arons~1998) postulate that the Lorenz factors of iron 
nuclei, which reach the region of the pulsar wind shock, should be
\begin{eqnarray}
\gamma_{\rm Fe}\approx 0.3Ze\Phi_{\rm open}/m_{\rm i}c^2
\label{eq20}
\end{eqnarray}
\noindent
where $m_{\rm i}$ and $Ze$ are the mass and charge of the iron nuclei, 
$c$ is the velocity of light, and $\Phi_{\rm open} = \sqrt{L_{\rm em}/c}$
is the total electric potential drop across the open magnetosphere.
These nuclei take significant part, $\chi$, of the total energy lost 
by the pulsar. In the model of Gallant \& Arons~(1994), the iron nuclei generate 
the Alfven waves in the down-stream region of the 
wind shock, which energy is resonantly transfered to positrons
present in the wind, as shown by particle-in-cell simulations by Hoshino et al.~(1992). 
As a result, the positrons obtain close to the power law spectrum with the
spectral index $\delta_1$ between 
$E_{\rm 1} = \gamma_{\rm Fe}m_{\rm e}c^2$ and $E_{\rm 2}\approx  
\gamma_{\rm Fe} A m_{\rm i}c^2/Z$ (see Gallant \& Arons~1994), where $m_{\rm e}$ and 
$m_{\rm i}$ are the electron and ion masses, respectively.
The spectrum is normalized to the conversion efficiency of energy from the iron nuclei 
to the positrons, $\xi$.  

For very fast pulsars the magnetic field in the shock region can be
high enough to limit the energies of positrons by synchrotron losses. These maximum 
energies of positrons can be estimated from comparison of the acceleration 
rate with the synchrotron energy loss rate, $dE/dt = \beta U_{\rm B}E^2$, where
constant $\beta = 4 c \sigma_{\rm T}/(3m_{\rm e}^2)$, $m_{\rm e}$ is the electron 
mass, $U_{\rm B} = B^2/8\pi$, and $B$ is the magnetic field strength in the 
acceleration region. The acceleration 
process discussed by Gallant \& Arons~(1994) occurs on a time scale 
corresponding to a few Larmor radii of particles with the maximum
allowed energies, i.e. at a rate
\begin{eqnarray}
\left({{dE}\over{dt}}\right)_{\rm acc} = {{c E}\over{2\pi r_{\rm g}\rho}}, 
\label{eq21b}
\end{eqnarray}
\noindent
where $\rho$ is the number of particles Larmor's radii, $r_{\rm g} = E/eB$.
The maximum energy of $\gamma$-ray photons, observed from the Crab Nebula,
allows to estimate the value of the parameter $\rho\approx 1$. 
Therefore, the maximum energies of positrons allowed by the synchrotron losses are 
\begin{eqnarray}
E_{\rm syn}^{\rm max}= \left({{8\pi e c}\over{\rho\beta B_{\rm sh}}}\right)^{1/2}
\approx 24(\rho B_{\rm sh})^{-1/2}~~~{\rm TeV}.
\label{eq21c}
\end{eqnarray}
\noindent
We assume that if $E_2 < E_{\rm syn}^{\rm max}$, then the cut-off in the positron 
spectrum is at $E_{\rm syn}^{\rm max}$. The rest of the energy transfered from nuclei 
to positrons with energies $E_{\rm syn}^{\rm max}$. We also postulate the injection of 
the $e^\pm$ pairs into the nebula with energies $E_1$, which are accelerated
in the pulsar wind. The energy contained in these $e^\pm$ pairs
is normalized to the part of the energy loss rate of the pulsar equal to 
$(1 - \chi)L_{\rm pul}$.

The free parameters of considered here model, $\chi$ and $\xi$, can be fixed for 
the specific PWNa by comparing the calculated synchrotron spectrum 
with the observations of the PWNa in the radio to X-ray energy range.

\section{The equilibrium spectra of relativistic particles inside the nebula}

In the previous section we defined the spectra of iron nuclei and leptons injected by 
the pulsar to the nebula as a function of time, measured from the pulsar formation. 
In order to determine the equilibrium spectra of these particles inside the nebula at 
a specific time we have to apply numerical approach. This is due to the fact that the
processes, which determine the escape and energy losses of injected particles, depend 
in a complicated way on the parameters of the nebula during its evolution. Below we 
calculate the equilibrium spectra of different types of 
nuclei (from disintegration of iron nuclei) and the equilibrium spectra of leptons
inside the nebula as a function of time.

\subsection{Spectra of nuclei}

\begin{figure}
  \vspace{12.truecm}
\includegraphics{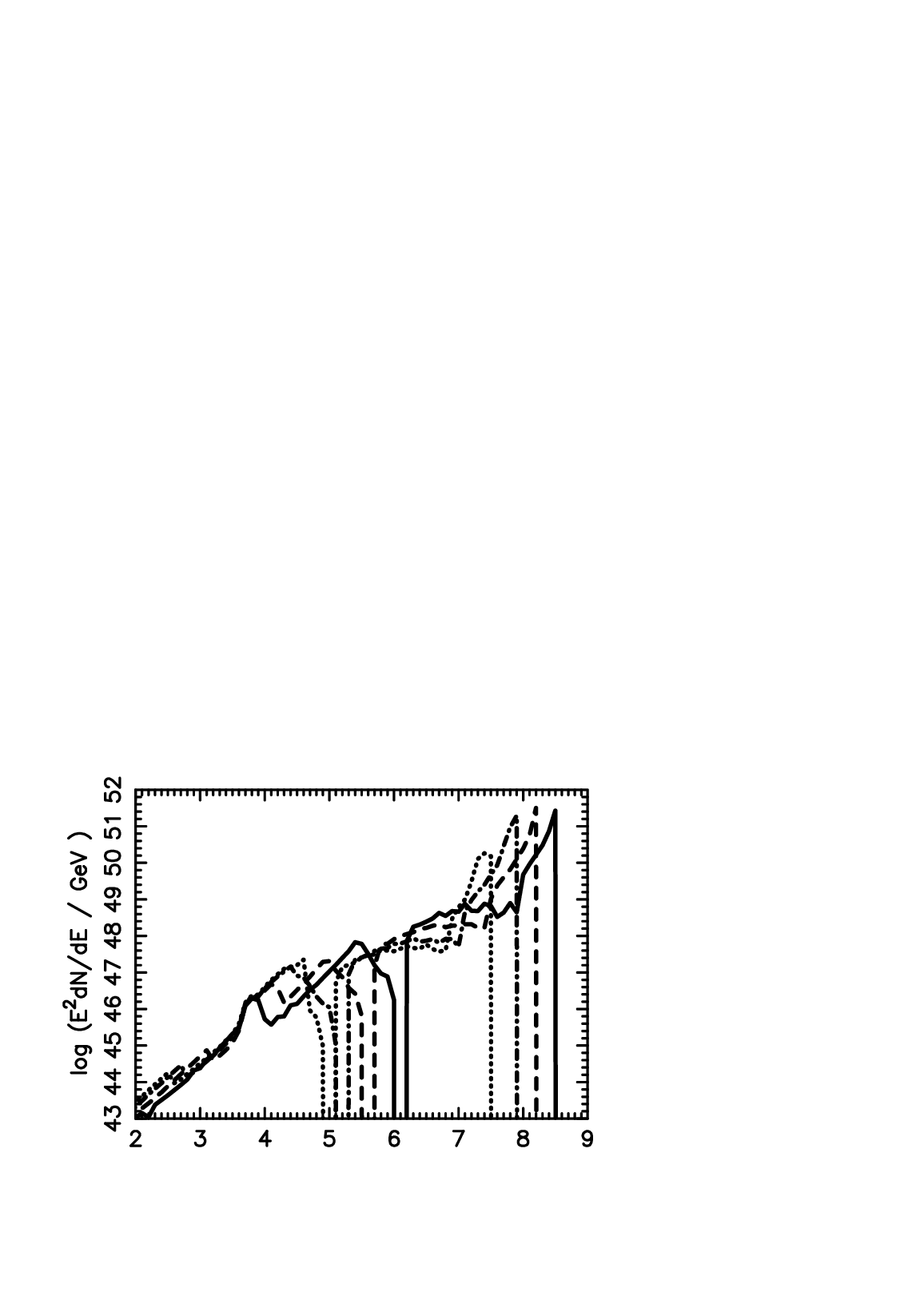}
\includegraphics{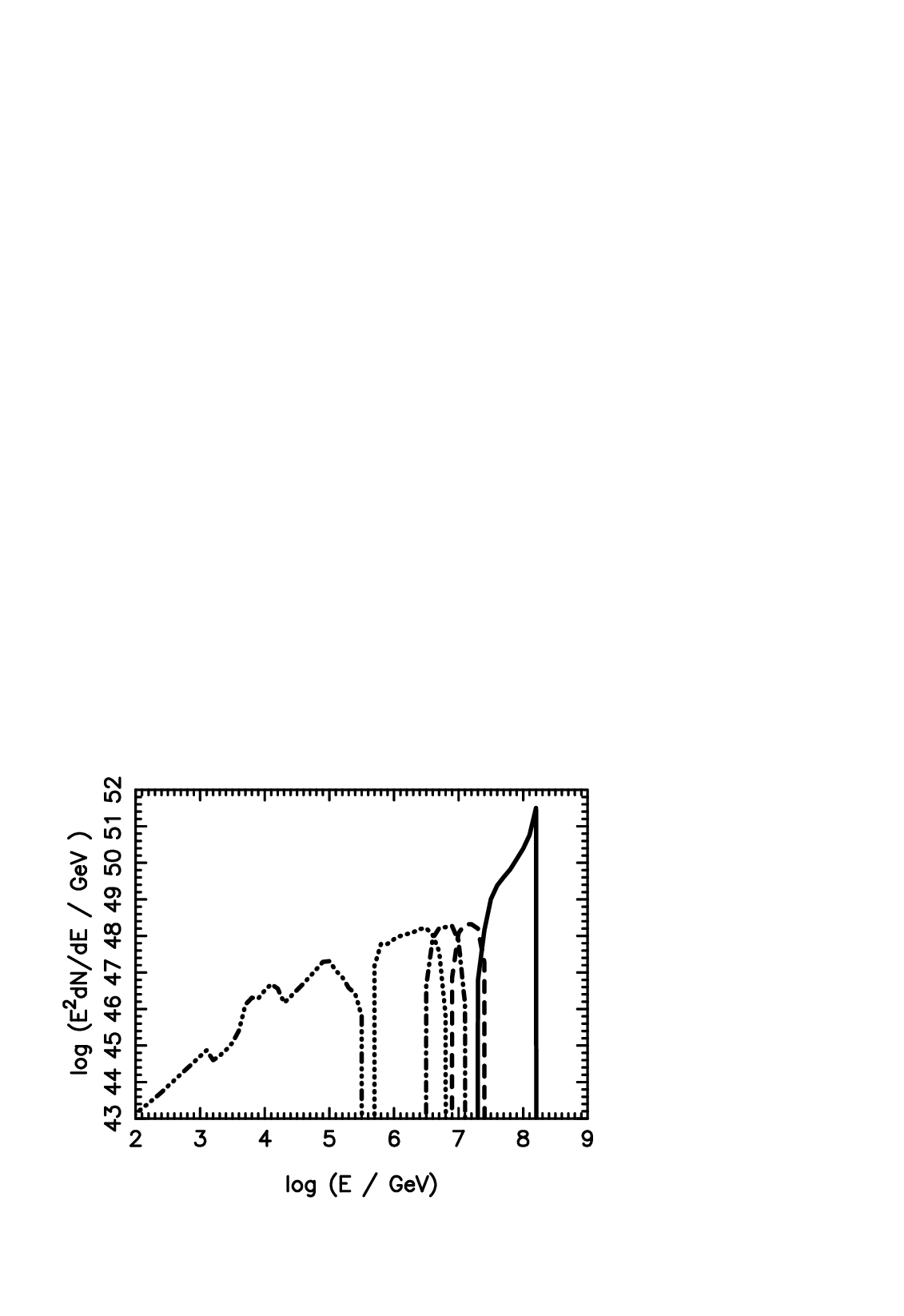}
  \caption{The equilibrium spectra of nuclei inside the nebula at specific time 
after the pulsar formation are shown in the upper figure for: $3\times 10^2$ yrs
(full curve), $10^3$ yrs (dashed), $3\times 10^3$ yrs (dot-dashed), and $10^4$ yrs 
(dotted). The initial parameters of the pulsar are $P_{\rm o} = 15$ ms and 
$B = 4\times 10^{12}$ G, the mass of the nebula $4 M_{\odot}$, and its expansion 
velocity 2000 km s$^{-1}$. The spectra of nuclei with different mass numbers, A, at 
$10^3$ yrs after the pulsar formation
are shown in the bottom figure: A = 1 (dot-dot-dot-dashed curve), 2-10 (dotted),
11-20 (dot-dashed), 21-40 (dashed), and 41-55 (full).}   
\label{fig2}
\end{figure}

\begin{figure}
  \vspace{6.truecm}
\includegraphics{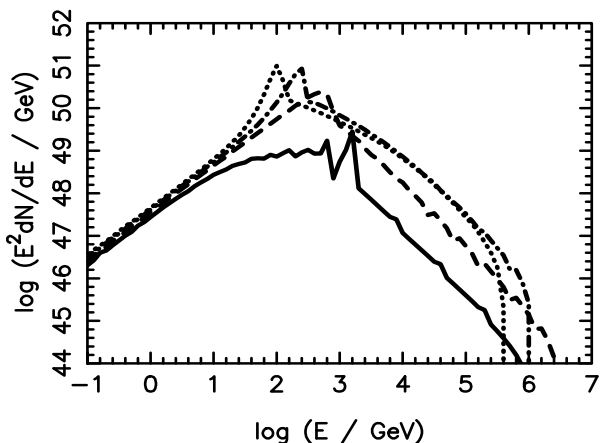}
  \caption{The equilibrium spectra of positrons at the time $3\times 10^2$ yrs 
(full curve), $10^3$ yrs (dashed), $3\times 10^3$ yrs (dot-dashed), and $10^4$ yrs 
(dotted) after the pulsar formation. Leptons are injected in two acceleration processes:
1) the positrons with the power law spectrum and spectral index 2.5, and 2) 
monoenergetic $e^\pm$, accelerated together with the
iron nuclei in the pulsar wind zone to the Lorentz factors $\gamma_{\rm Fe}$.   
It is assumed that nuclei take $\chi = 0.85$ of the rotational energy lost by the pulsar
and accelerates positrons with the efficiency $\xi = 0.5$.
The rest of available energy, $1 - \chi = 0.15$, is transfered to the monoenergetic 
$e^\pm$ pairs.} 
\label{fig3}
\end{figure}

The nuclei injected into the nebula can lose energy as a result of the acceleration of 
positrons, collisions with the matter of the nebula, and due to adiabatic expansion of 
the nebula. At the early stage, when the nebula is relatively dense, nuclei also fragment 
to lighter products due to the collisions with the matter. The most 
energetic nuclei can also diffuse out of the nebula. 
All these processes, adiabatic and collisional losses, fragmentation, escape,
have been recently studied in detail by Bednarek \& 
Protheroe~(2002) with the purpose to calculate the contribution of pulsar accelerated
nuclei to the cosmic rays in the Galaxy. In order to
calculate the equilibrium spectra of different types of nuclei inside the nebula at 
a specific time, we follow the method adopted in Bednarek \& 
Protheroe~(2002), and described in detail in Sect. 4.2 and 4.3 of that paper.
The only difference concerns the model for the expansion of the nebula which in the 
present paper include the effects of energy supplied by the pulsar to 
the nebula (Sect.~2).

However, in contrary to Bednarek \& Protheroe~(2002), who assumed that nuclei are 
accelerated only by the electric field of the outer gap in the pulsar magnetosphere 
(Cheng, Ho \& Ruderman~1986), we make use of the model proposed by Gallant \& Arons~(1994) 
(see details in section~3 in that paper), in which the nuclei can reach energies 
corresponding to 
the significant part of the electric field drop through the pulsar polar cap.  
Therefore, the iron nuclei, extracted from the neutron star surface, are: (1) at 
first accelerated in the outer gap where they can lose only a small number of nucleons in 
collisions with the radiation field of the outer gap (see section 4.1 in Bednarek \& 
Protheroe~2002 and Bednarek \& Protheroe 1997) and, (2) farther accelerated 
in the pulsar wind zone to the Lorentz factors $\gamma_{\rm Fe}$ (Eq.~8),  
determined by the pulsar parameters. 
 
As an example, in Fig. 2 we show the equilibrium spectra of nuclei inside the nebula for 
the case of the pulsar with initial 
parameters $P_{\rm o} = 15$ ms and $B = 4\times 10^{12}$ G, the mass of the nebula
4 M$_{\odot}$, and its expansion velocity 2000 km s$^{-1}$ (as expected for the Crab 
pulsar).
The upper figure shows the spectra of nuclei at different time after the explosion of
supernova. As expected the maximum energies and the numbers of relativistic nuclei 
inside the nebula decrease with time due to the adiabatic energy losses and escape of 
the most energetic nuclei from the nebula. However the spectra of nuclei at lower
energies (lighter nuclei) do not change significantly.
The bottom figure shows the spectra of nuclei with different mass numbers at $10^3$ years
after the supernova explosion. The spectrum of nuclei is dominated by 
the heavy elements from the iron group. These nuclei have been mainly injected into the 
nebula when it becomes transparent for disintegration of nuclei. 
This typically happens at several to a few tens of years after the supernova explosion.

\subsection{Spectra of leptons}

\begin{figure*}
  \vspace{6.truecm}
\includegraphics{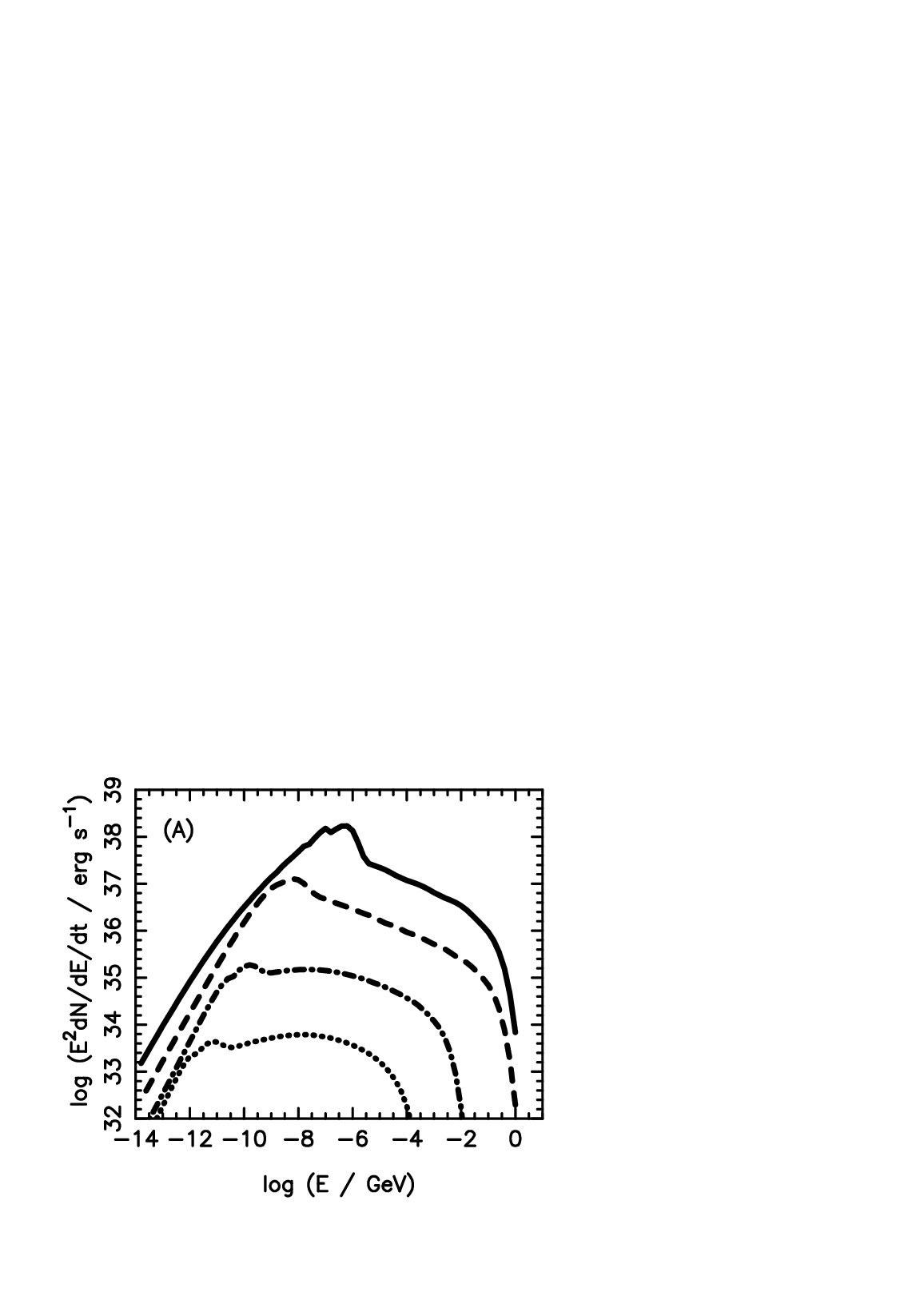}
\includegraphics{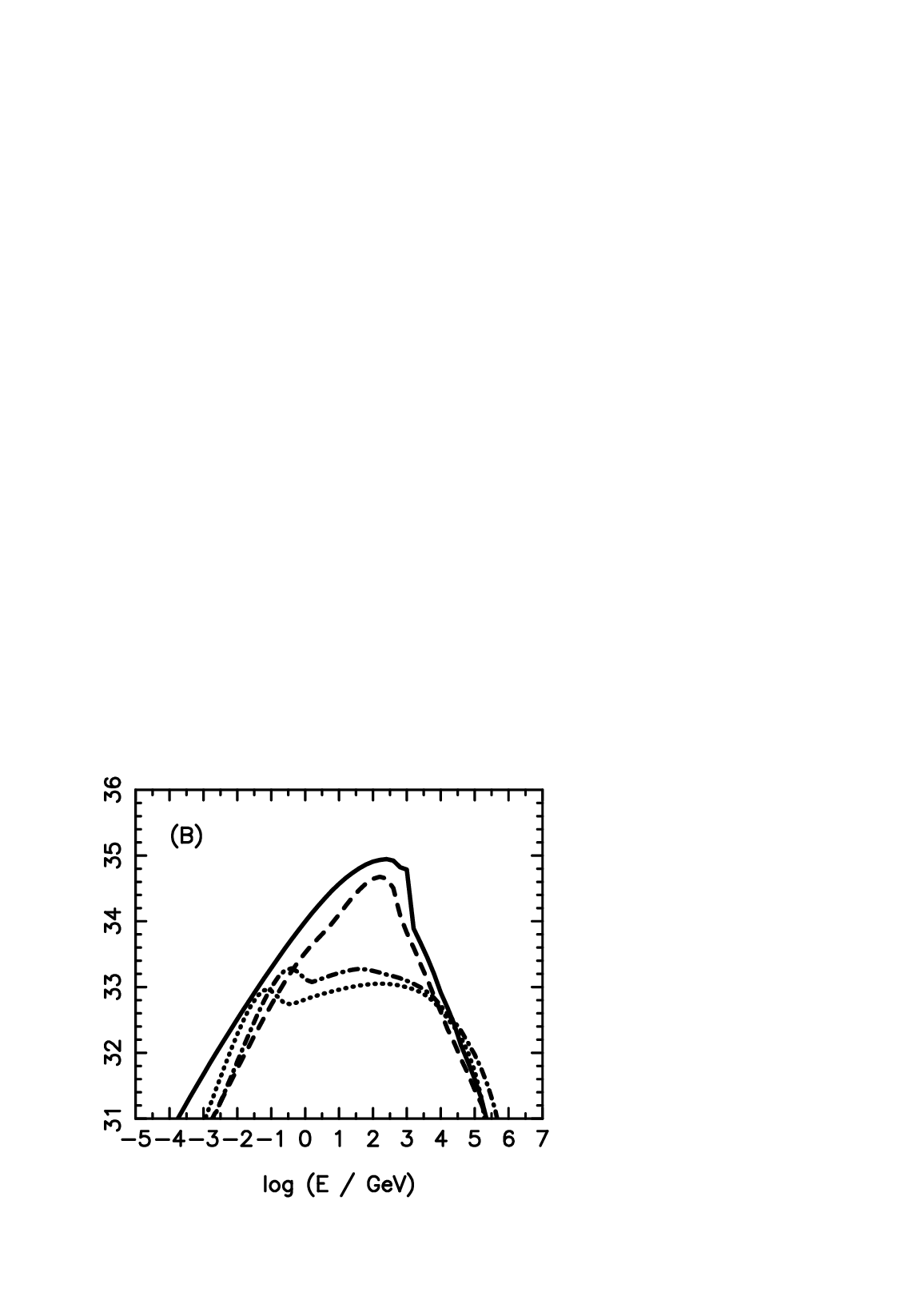}
\includegraphics{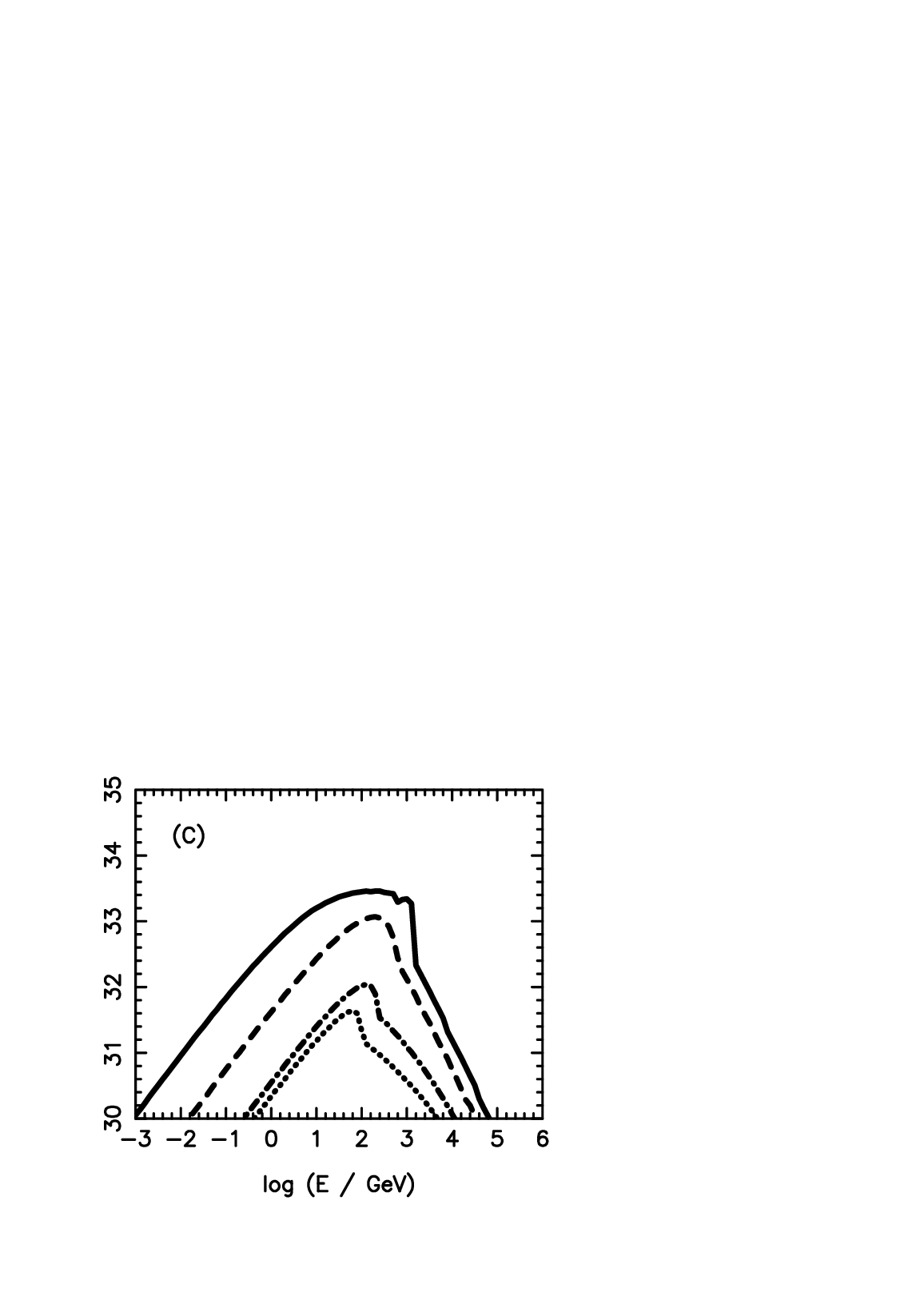}
  \caption{Photon spectra produced by leptons in synchrotron (A), inverse 
Compton (B), and bremsstrahlung (C) processes, at the time 
$3\times 10^2$ yrs (full curves), $10^3$ yrs (dashed), $3\times 10^3$ yrs (dot-dashed), 
and $10^4$ yrs (dotted) after the supernova explosion. Positrons are injected into the 
nebula with the power law spectrum and spectral index 2.5
between $E_1$ and $E_2$ (see text), normalized to the pulsar 
energy loss rate with $\chi = 0.85$, and the acceleration efficiency $\xi = 0.5$.
The monoenergetic $e^\pm$ pairs are injected with energies $E_1$.}   
\label{fig4}
\end{figure*}

Positrons injected into the medium of expending supernova 
remnant suffer energy losses mainly on radiation processes, bremsstrahlung, synchrotron, 
and the inverse Compton, and due to the expansion of the nebula. 
The rate of their energy losses can be described by 
\begin{eqnarray}
-{{dE}\over{dt}} = (\alpha_1 + \alpha_2)E + (\beta_1 + \beta_2) E^2~~~{\rm GeV~s^{-1}},
\label{eq22}
\end{eqnarray}
\noindent
where $\alpha_1\approx m_{\rm p} N/(m_{\rm e} X)\approx 7.8\times 10^{-16} N$ s$^{-1}$ 
describes the bremsstrahlung losses, where $m_{\rm p}$ is the proton mass,
X is the radiation length and $N$ is the density of the medium in
cm$^{-3}$; $\alpha_2 = V_{\rm Neb}(t)/R_{\rm Neb}(t)$ describes the adiabatic losses
due to the expansion of the nebula (Longair 1981); 
$\beta_1\approx 4c\sigma_{\rm T}U_{\rm B}/(m_{\rm e}^2c^5) \approx 2.55\times 
10^{-6}B^2$ GeV$^{-1}$ s$^{-1}$ the synchrotron energy losses, where $\sigma_{\rm T}$ is 
the Thomson cross section, $U_{\rm B}$ is the magnetic field energy density and $B$ is 
the magnetic field in G; and $\beta_2\approx 
4c\sigma_{\rm T}U_{\rm rad}/(m_{\rm e}^2)\approx 1.05\times 
10^{-7}U_{\rm rad}$ GeV$^{-1}$ s$^{-1}$ the ICS losses in the Thomson regime. 
$U_{\rm rad}$ is the energy density of different types of soft radiation inside the 
nebula, i.e. the synchrotron radiation created by leptons in the magnetic field of the 
nebula, the MBR, and the infrared photons produced by the dust inside the nebula. 
The energy losses of leptons on the ICS in the Klein-Nishina regime 
can be safely neglected in respect to the synchrotron energy losses. The density of
synchrotron radiation depends on the spectrum of positrons which is in turn determined 
by their energy losses at the earlier phase of expansion of the nebula. 

The coefficients, $\alpha_1, \alpha_2, \beta_1$, and $\beta_2$, depend on time 
in a complicated way due to the changing conditions in the expending nebula (magnetic
field, density of matter and radiation). Therefore, Eq.~(11) can not be integrated 
analytically. In order to determine the energies of leptons, $E$, inside the 
nebula at a specific time, $t_{\rm obs}$,
which have been injected with energies, $E_{\rm o}$, at an earlier time, t', 
we use the numerical approach. The equilibrium spectrum of leptons is then obtained  
by integrating their injection spectra over the age of the nebula 
\begin{eqnarray}
{{dN(t_{\rm obs})}\over{dE}} = \int_{0}^{t_{\rm obs}} J(t')
 {{dN}\over{dE_{\rm o}}dt}dt,
\label{eq24}
\end{eqnarray}
\noindent
where $dN/dE_{\rm o}dt$ is the injection spectrum of positrons at the time t, defined in
Sect.~3, $t' = t_{\rm obs} - t$, and the jacobian $J(t') = E_{\rm o}/E$.

The example spectra of leptons inside the nebula at the specific time after explosion
of supernova are shown in Fig.~3, assuming that leptons are injected into the 
nebula with the two component spectrum: 1) the power law spectrum normalized to the 
pulsar energy loss rate, $\chi = 0.85$, and efficiency of energy conversion from nuclei 
into positrons, $\xi = 0.5$, and the spectral index of the positrons equal to 
$\alpha = 2.5$; 2) The monoenergetic $e^\pm$ pairs accelerated in the pulsar wind zone 
to energies $E_1$. The density of leptons inside the nebula 
is lower when the nebula is younger in spite of the higher rate of injected leptons.
This is due to the much higher efficiency of energy loss processes of 
positrons in young nebula when the magnetic field and the radiation energy densities are
much larger. That's why, the equilibrium spectra of positrons cuts-off at lower energies 
when the nebula is young. However, at later stage of the nebula, the cut-off in the 
spectrum of leptons occurs again at the lower
energies due to the lower energies of injected leptons. The maximum in the equilibrium 
spectrum of leptons shifts to higher energies at the early stage, and then drops to lower 
energies for the nebulae with the age larger than $\sim 10^3$ years.

\section{Production of radiation inside the Nebula}

The knowledge on the equilibrium spectra of relativistic particles (leptons and nuclei) 
as a function of time after supernova explosion allows us to calculate the photon 
spectra produced inside the nebula by these particles in different radiation processes. 
Leptons produce photons mainly on synchrotron, bremsstrahlung, and ICS processes. 
Nuclei produce $\gamma$-rays in collisions with the matter via decay of neutral pions. 
We neglect the radiation produced by the secondary $e^\pm$ pairs from decay of charged 
pions, since their contribution to the equilibrium spectrum of leptons inside the nebula 
is negligible in respect to the contribution from primary positrons, accelerated directly 
in resonant interactions with heavy nuclei and in the pulsar wind zone.

\subsection{Radiation from leptons}

The conditions in the expanding nebula, i.e. the magnetic and radiation fields and the
density of matter, change significantly with time in a different manner. Thus, the 
relative importance of specific radiation processes has to change as well. At the early 
stage of expansion of the nebula, the synchrotron energy losses of leptons dominate over 
the ICS and the bremsstrahlung energy losses. Therefore, most of the energy of leptons is 
radiated in the low energy range. When the nebula becomes older, the energy density 
of the synchrotron radiation inside the nebula decreases but the energy density of 
microwave background radiation (MBR) remains constant. Therefore, relative
importance of the ICS losses increases in respect to the synchrotron energy losses.
The energy losses of leptons on synchrotron process and bremsstrahlung process
can be simply compared for known parameters of the nebula at a specific time.
The bremsstrahlung process dominates over the synchrotron process for leptons with 
energies,
\begin{eqnarray}
E < \alpha_1/\beta_1\approx  4.1\times 10^{-10} N B^{-2} ~{\rm GeV}.
\label{eq25}
\end{eqnarray}
For example, if the age of the nebula is $10^3$ yrs (the Crab Nebula),
then the density of matter is $N \sim 10$ cm$^{-3}$ and the magnetic field 
$B\sim 10^{-4}$ G. For these parameters, the synchrotron energy losses of leptons 
dominate over their bremsstrahlung energy losses for energies as low as 
$\sim 0.4$ GeV. Therefore, if there exists large population of hundred MeV leptons
inside the Crab Nebula, they should radiate bremsstrahlung photons with the MeV energies. 

In order to determine precise contribution of these three processes, 
we calculate the photon spectra produced by leptons with the equilibrium spectrum, 
obtained as shown in the previous section. As an example, in Fig. 4 we show the photon 
spectra from these three process at a specific time after supernova explosion for the 
equilibrium spectra of leptons obtained for the initial parameters of 
the pulsar and nebula as defined in Fig. 3. It is assumed that 
leptons produce the IC $\gamma$-rays by scattering
the synchrotron and MBR photons. Note that the synchrotron spectra from the nebula 
strongly decrease with time, and the contribution of the bremsstrahlung spectra to the
$\gamma$-ray energy range is always at least an order of magnitude lower than the 
contribution of the IC spectra.

\subsection{Gamma-rays from hadrons}

The nuclei with the equilibrium spectra interact with the matter of the nebula and 
produce $\gamma$-rays via decay of pions. We calculate $\gamma$-ray spectra 
from hadronic interactions applying the scaling break model proposed by Wdowczyk \& 
Wolfendale~(1987).
As an example, we show in Fig.~5 the spectra of $\gamma$-rays produced 
at different time after the supernova explosion. The initial parameters of the
pulsar and the supernova are taken as in the example shown for the photon
spectra produced by leptons.
The intensities of the $\gamma$-ray spectra decrease with the age of the nebula 
due to the lower densities of matter and relativistic nuclei inside the nebula.
This decrease is much faster than observed in the case of the $\gamma$-ray spectra 
produced by leptons in ICS process. Therefore, the contribution of $\gamma$-rays
from hadronic processes is relatively less important for older nebulae. The $\gamma$-ray 
spectra from hadronic collisions shift to lower energies due to the adiabatic energy 
losses of nuclei, more efficient escape of higher energy nuclei from the nebula, and 
lower energies of freshly injected nuclei inside the nebula by older pulsars.

\begin{figure}
  \vspace{6.truecm}
\includegraphics{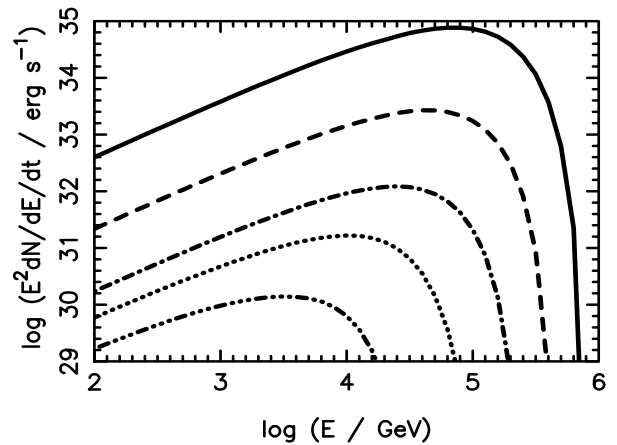}
  \caption{Spectra of gamma-rays produced in the interactions of nuclei,
with the equilibrium spectra shown in Fig.~2, with the matter of the nebula 
at a specific time after the pulsar formation: 
$3\times 10^2$ yrs (full curve), $10^3$ yrs (dashed), $3\times 10^3$ yrs (dot-dashed), 
$10^4$ yrs (dotted), and $3\times 10^4$ yrs (dot-dot-dot-dashed).
The initial parameters of the pulsar and the nebula are as in Figs.~2 and 3.}   
\label{fig5}
\end{figure}
\section{Application to specific supernova remnants}

The radiation model, considered above, is at first confronted with the observations of 
the best investigated WPNa, i.e. the Crab Nebula. Our strategy is to fix some free 
parameters of the model based on the comparison of the observed radio up to the TeV 
$\gamma$-ray spectrum with our calculations of the synchrotron, ICS, and $\gamma$-ray 
spectrum from decay of pions. Trying to keep constant as many as possible parameters 
derived for the Crab Nebula, we calculate the high energy photon spectra for the cases 
of other PWNe, which were also reported at the TeV $\gamma$-rays, i.e. Vela pulsar and 
the nebula around PSR 1706-44.  
Interestingly, the observed parameters of these pulsars and nebulae can be 
obtained assuming these same initial parameters as found for the Crab pulsar,  
i.e. it seems that these two pulsars and nebulae are on these same evolutionary 
path as the Crab pulsar and nebula. Using the experience obtained from modelling the 
nebulae with known $\gamma$-ray emission, we use this model to predict the level of TeV 
$\gamma$-ray emission from the nebulae discovered around other energetic pulsars: 
J0205+6449 in the nebula 3C 58, and PSR 1951+32 in the nebula CTB80.

\subsection{The Crab Nebula}

\begin{figure}
  \vspace{10.5truecm}
\includegraphics{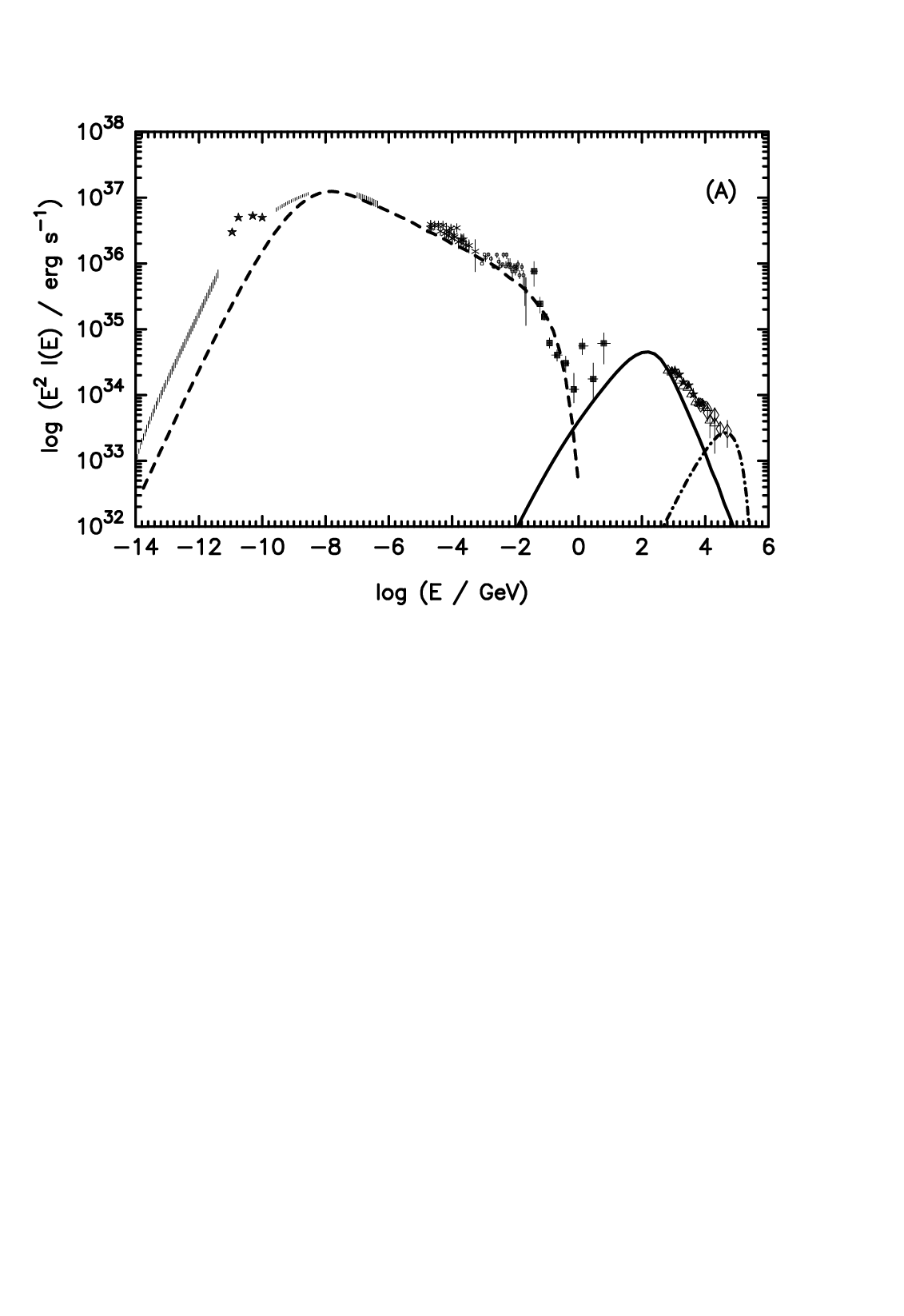}
\includegraphics{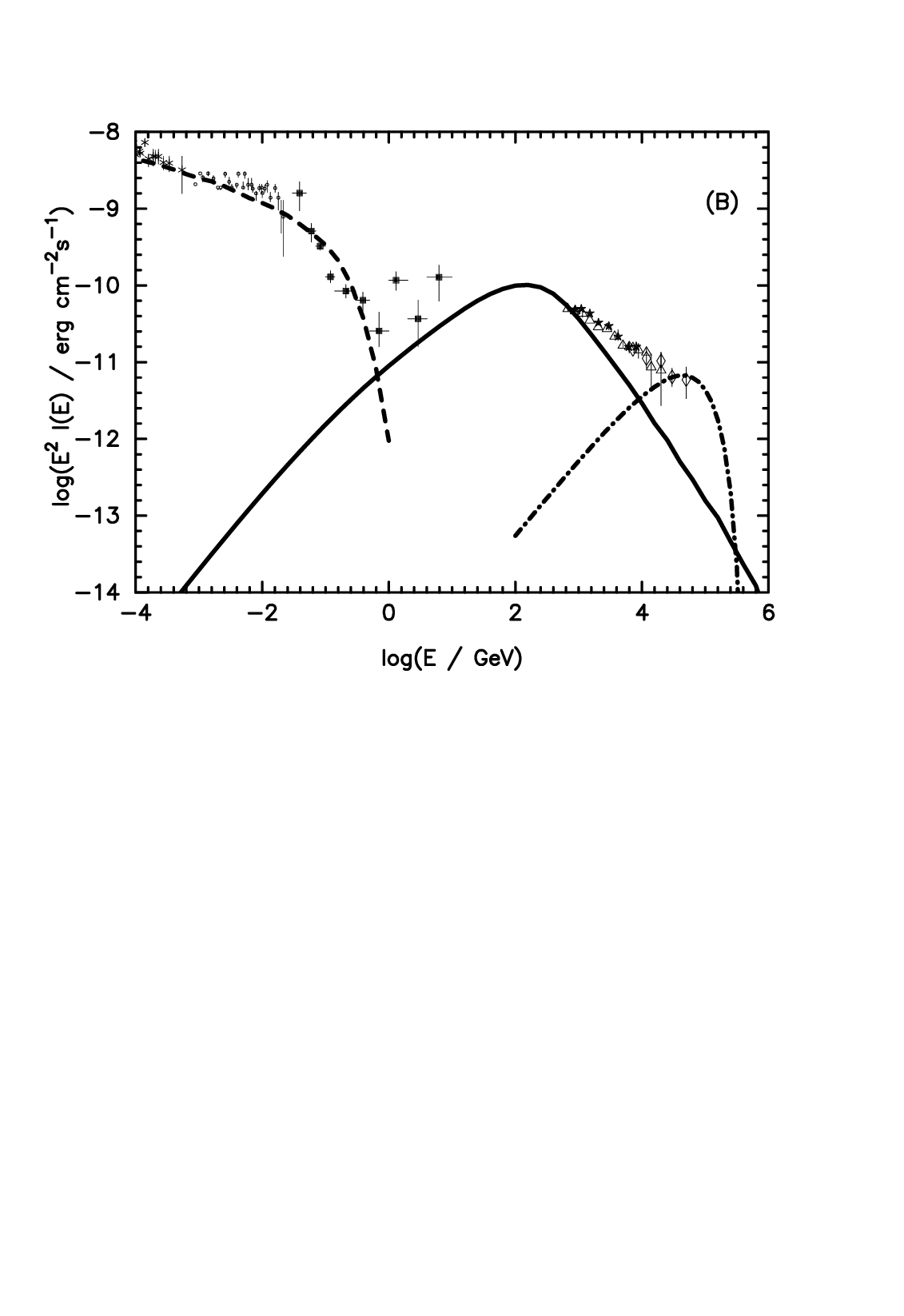}
  \caption{The multiwavelength spectrum of the Crab Nebula compared with the 
spectrum calculated in terms of the hadronic-leptonic model (A). 
The details of the Crab Nebula spectrum in the $\gamma$-ray energy range (B).
Leptons are injected into the nebula with the power law spectrum (positrons 
accelerated in the interactions with iron nuclei) and spectral index 
$\alpha = 2.5$ betwen $E_1$ and $E_2$. It is assumed that the iron nuclei 
take $\chi = 0.85$ of the total rotational energy lost by the pulsar. A part of 
the energy of the iron nuclei, $\xi = 0.5$, is converted into relativistic positrons. 
The rest of the pulsar rotational energy, $1 - \chi$, is curried out by the 
monoenergetic $e^\pm$ pairs with energies $E_1$. The synchrotron and
inverse Compton spectra (comptonization of synchrotron, MBR, and infrared photons),
produced by leptons inside the nebula, are shown by the dashed and full curves, 
respectively. The $\gamma$-ray spectrum from decay of pions, produced by hadrons 
inside the nebula, is shown by the dot-dashed curve. The other parameters of the pulsar
and nebula are given in the text.}   
\label{fig6}
\end{figure}

The Crab Nebula is the only one WPNa which has been detected in the 
all energies of the electromagnetic spectrum, starting from the radio up to a few tens 
of TeV $\gamma$-rays. The Crab Nebula photon spectrum consists of two broad bumps: first 
one, from radio up to MeV $\gamma$-rays, with the spectrum described by simple power law 
with the index $\sim 2.25$ from optical to MeV $\gamma$-rays; and the second one, from a 
hundred MeV up to $\sim 50$ TeV, described also by the power law spectrum with the 
spectral index $\sim 2.5$ between $0.1 - 50$ TeV (see Fig.~6). 

As in other works, we interpret the first broad bump in the Crab 
Nebula spectrum as synchrotron emission produced by leptons inside the nebula. 
The second bump is interpreted as the ICS of soft photons inside the nebula by these
same population of leptons and by self-consistently calculated $\gamma$-ray emission
from decay of charged pions, produced in collisions of nuclei with the matter of the 
nebula. In order to perform calculations we have to fix at first the initial parameters 
of the Crab pulsar and nebula. The observed period, 33.4 ms, and the age of the
nebula, $\sim 950$ yr, are consistent with the Crab pulsar initial period of 15 ms 
and its surface magnetic field of $3.8\times 10^{12}$ G. The observed parameters of the 
Crab Nebula, the radius of 2-3 pc for the distance 1830 pc, the expansion velocity 
$\sim 2000$ km s$^{-1}$, and the mass of the Crab Nebula equal to $\sim 4 M_{\odot}$ 
(Davidson \& Fesen~1985),
are consistent with the present age of the pulsar. We assume that the initial expansion
velocity of the bulk matter of the nebula is very close to the observed velocity since
the nebula is still in the free expansion phase for the typical density of the 
surrounding medium 0.3 cm$^{-3}$.

For the above initial parameters of the Crab pulsar and nebula we have performed 
a sequence of calculations of the expected broad range spectra in order to obtain good 
fit to the observed spectrum. As described in Sect.~3, the injection spectrum of leptons
consists of two power laws. Above the energy $E_1$, the spectral index has to be 
$\delta_1 = 2.5$, in order to be consistent with the observed synchrotron spectrum with 
the spectral index in the broad energy range equal to $\sim 2.25$. 
If the iron nuclei take, $\chi = 0.85$, of the total energy lost by 
the Crab pulsar, then the calculated synchrotron emission is consistent with 
the observed emission (see dashed curve in Fig.~6), for the efficiency of energy 
conversion from the iron nuclei to the positrons equal to $\xi = 0.5$. We assume, 
moreover, that the rest of energy lost by the pulsar, $1 - \chi$, is curried out by the
monoenergetic $e^\pm$ pairs with the Lorentz factors of the iron nuclei and energies
$E_1$. For these parameters we can fit the observed synchrotron spectrum in the broad
energy range. However, the calculations do not give correct description of the lowest 
energy part of the spectrum (below optical). This is due to the assumption that all 
$e^\pm$ pairs are accelerated in the pulsar wind to these same monoenergetic
energies equal to $E_1$. In a more realistic model, the pairs should be accelerated in 
the pulsar wind to different maximum energies, depending on their injection from 
different parts of the light cylinder. As a consequence, the $e^\pm$ pairs should 
be injected into the nebula with some spectrum which shape is not precisely known
by the theory.

We also calculate the ICS spectrum
from the Crab Nebula, applying that the soft photon field for leptons is created by the 
synchrotron photons produced by this same leptons, microwave background radiation (MBR),
and the thermal infrared photons, emitted by the dust inside the nebula with the energy
density equal to the energy density of the MBR and the temperature 100 K (see Aharonian \&
Atoyan~1995). The reasonable consistency with the observed level of $\gamma$-ray emission 
from the Crab nebula at $\sim 1$ TeV is obtained if we assume that the energy density of 
synchrotron photons is $\sim 20$ times higher in the region where the most of the IC 
photons are produced. In another words
the effective volume  of synchrotron emission has to be $\sim 20$ times lower than the 
whole volume of the nebula. In fact, leptons are not injected into the nebula 
symmetrically but at some range of angles close to the equatorial plane of the rotational 
axis of the pulsar as suggested by the observations of the inner X-ray torus. 

The calculated ICS spectrum is too steep and does not fit 
correctly the $\gamma$-ray spectrum observed from the Crab Nebula above a few TeV
(see full curve in Fig.~6). However, calculated selfconsistently the spectrum of 
$\gamma$-ray photons from decay of pions produced by heavy nuclei, which accumulate 
inside the nebula, fits very well the $\gamma$-ray emission at $\sim 10$ TeV 
(see dot-dashed curve in Fig.~6). Note that $\gamma$-ray emission from nuclei does not 
depend on any additional free parameters since in the considered here model the numbers 
of relativistic nuclei and leptons are strictly connected.

\subsection{The Vela Supernova Remnant}

\begin{figure}
  \vspace{5.6truecm}
\includegraphics{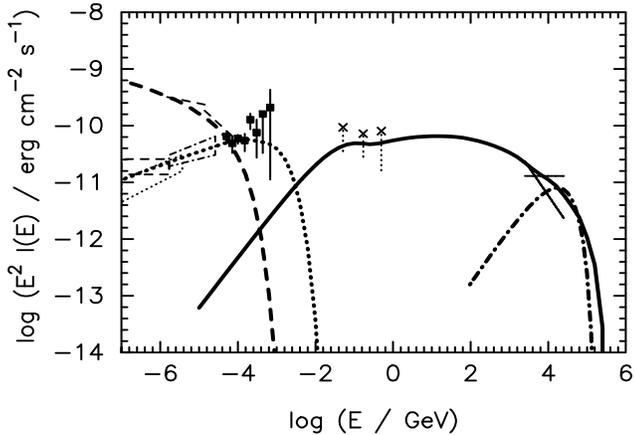}
  \caption{The high energy spectrum observed from the compact nebula around the Vela 
pulsar and from the Vela SNR (see text for details) is compared with the spectrum 
calculated in terms of the discussed hadronic-leptonic model. Relativistic leptons 
produce synchrotron spectrum during propagation through the compact nebula with the 
strong magnetic field (dotted curve) and after that, in the extended Vela SNR 
(dashed curve). The IC spectrum, produced by leptons in the Vela SNR due to the 
comptonization of the MBR, is shown by the full curve. The $\gamma$-ray spectrum from 
decay of pions, produced by nuclei inside the Vela SNR, is shown by the dot-dashed 
curve. The initial parameters of the pulsar and supernova are as in the case of the 
Crab Nebula (accept the magnetic field of the pulsar) and the real age of the nebula 
equal to 6000 yrs.}   
\label{fig7}
\end{figure}

The observations of the Vela pulsar performed with the large field and small field
detectors show two different spectral shapes in the X-ray energy range. 
The large field instruments, OSO-8  (Pravdo et al.~1978) 
and HEAO A-4 (Levine et al.~1984), observe steep emission from the whole Vela 
Supernova Remnant (SNR), which dominates at energies below $\sim 10$ keV.
On the other hand, the small field detectors, {\it Einstein} (Harnden et al.~1985), 
{\it EXOSAT} (\"Ogelman \& Zimmermann~1989),
{\it ROSAT} (\"Ogelmanan, Finley \& Zimmermann~1993), and {\it Birmingham Spacelab 2}
(Willmore et al.~1992), observe flatter emission with the differential 
spectrum and average spectral index  $\sim 1.7$. This spectrum is consistent
with the observations of Vela supernova remnant by OSSE between 44 keV and 370 keV
(De Jager et al.~1996). The discrepancy 
between results reported by the large and small field detectors can be explained by
assuming that the X-ray spectrum of the Vela compact nebula extends to higher energies 
and is flatter than the emission observed from the Vela supernova remnant. 
The unpulsed TeV $\gamma$-ray emission has been also reported from the direction of the 
Vela pulsar at energies $> 2.5$ TeV (Yoshikoshi et al.~1997). At lower energies,
only the upper limits are available in the EGRET energy range (Kanbach et al.~1994) and
above 300 GeV (Chadwick et al.~2000).

The Vela pulsar is surrounded by the compact non-thermal nebula with the radius of 
$\sim 7'$ and the extended Vela SNR with the radius of $\sim 3.5^{\rm o}$. 
The observed parameters of the pulsar allow us to estimate its surface magnetic field, 
$4.5\times 10^{12}$ G, and the characteristic age of 11.300 yrs. We model the Vela SNR
assuming that the initial parameters of this object were similar to the Crab Nebula, 
i.e. the initial pulsar period 15 ms, the mass of Vela SNR 4 M$_\odot$ and the expansion 
velocity 2000 km s$^{-1}$. The distance to the Vela pulsar 
is taken $\sim 300$ pc (Caraveo et al.~2001), although the older literature suggested the 
value of 500 pc (e.g. Cha, Sembach \& Danks~1999). For these parameters of the pulsar 
and nebula, we estimate its real age $\sim 6000$ yrs, from the consistency with the 
observed dimension of the nebula and the observed period of the Vela pulsar. 

We consider the radiation processes around the Vela pulsar as a two stage
process. At first leptons are injected into the inner nebula with a
relatively small size, $\sim 0.15$ pc. They move through the Vela compact nebula with 
high velocity (about one third of the velocity of light) losing only part of their 
initial energies. The average magnetic field in the Vela WPN is estimated on 
$4\times 10^{-5}$ G. After that, leptons are injected into the large scale
Vela SNR, where the magnetic field is close to that in the interstellar medium, 
$\sim 5\times 10^{-6}$ G. The model with above magnetic field strengths can explain the 
different spectra observed by the large and small field X-ray detectors.

In order to fit the expected X-ray and $\gamma$-ray spectra from the Vela WPNa
by the synchrotron and IC spectra calculated in terms of such a model,
we assume that the iron nuclei take similar part of the energy lost by the 
pulsar as in the case of the Crab pulsar, i.e. $\chi = 0.85$. However, to obtain the
correct level of synchrotron emission from the inner and the outer nebulae, we
were forced to assume that the average acceleration efficiency of positrons by the 
nuclei is much lower than in the case of the Crab pulsar and equal to $\xi = 0.07$.
In Fig.~\ref{fig7}, the results of calculations of the synchrotron spectra and the
IC spectrum are compared with the 
observations of the Vela PWNa in the X-rays, with the upper limits
from the EGRET observations (Kanbach et al.~1994), and with the TeV $\gamma$-ray 
energy range (Yoshikoshi et al.~1997).
The emission from the extended Vela SNR fits well to the total 
spectrum from the Vela nebula below $\sim 10$ keV and is responsible for the 
$\gamma$-ray emission observed in the TeV energy range. Note that the IC spectra are 
produced mainly due to the scattering of the MBR. The synchrotron photons 
can be neglected for nebulae with the age of the Vela pulsar.
On the other hand, the 
emission from the compact nebula dominates in the hard X-rays (between $\sim 10$ keV and 
$\sim 1$ MeV). The contribution of leptons from the compact nebula to the high energy 
$\gamma$-ray spectrum is negligible since we assumed that they move relatively
fast through the inner nebula. 
We have also calculated the $\gamma$-ray spectrum from decay of pions, produced 
by heavy nuclei injected by the pulsar (dot-dashed curve in Fig.~7). Their contribution
to the total $\gamma$-ray spectrum observed from the Vela SNR above $\sim 10$ TeV 
is still comparable to the IC spectrum, but is relatively lower than in the case of 
the Crab Nebula.  

\subsection{The Nebula around PSR 1706-44}

\begin{figure}
  \vspace{5.6truecm}
\includegraphics{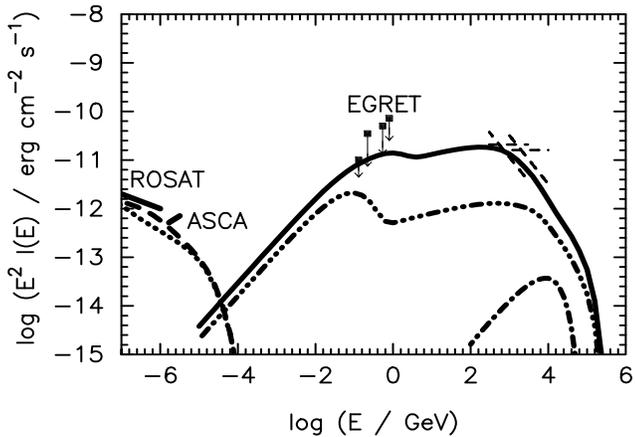}
  \caption{The high energy spectrum from the nebula (G 343.1-2.3) around the pulsar
PSR 1706-44 (see text for details) is
compared with the synchrotron spectrum and the IC spectrum,
produced by positrons accelerated in interactions with nuclei and $e^\pm$ pairs 
accelerated in the pulsar wind zone. 
The synchrotron and $\gamma$-ray spectra calculated under the assumption that 
leptons comptonize only the MBR are shown by the dashed and dot-dot-dot-dashed curves, 
respectively. These spectra obtained with the additional strong infrared field 
inside the nebula with the temperature 20 K are shown 
by the dotted and full curves, respectively. The initial parameters of the pulsar and 
the supernova are as in the case of the Crab Nebula. The surface magnetic field of the 
pulsar is $3.1\times 10^{12}$ G, and its age, 16.000 yrs.
The $\gamma$-ray spectrum from decay of pions is shown by the dot-dashed curve.
}   
\label{fig8}
\end{figure}

The pulsar PSR 1706-44 has been associated with the
supernova remnant SNR G 343.1-2.3 on the basis of radio observations (McAdam, Osborne \& 
Parkinson 1993; Dodson \& Golap~2002), although there are also some distance 
inconsistencies  between these two objects (Frail, Goss \& Whiteoak 1994). 
PSR 1706-44 shows close similarities to the 
Vela pulsar and its SNR since these pulsars have similar characteristic ages and present 
periods. PSR 1706-44 is also immersed in a 
very compact nebula with a radius $\sim 0.1-0.3$ pc (Becker, Brazier \& Trumper 1995). 
The spectrum of this nebula in the energy range 0.1 - 2.4 keV has the differential photon 
index $2.4\pm 0.6$, and the spectrum reported in the higher energy 
range (0.7 - 10 keV) has the index 1.7 (ROSAT and ASCA observations, Finley et al. 1998).
The ROSAT upper limit on the flux from the supernova remnant SNR G 343.1-2.3 is 
$9\times 10^{-13}$ erg cm$^{-2}$ s$^{-1}$, assuming the spectrum with the differential 
photon index equal to 2 (Becker et al.~1995).
Thus the spectral features of the emission from the compact and extended nebulae are
similar to this observed in the nebulae surrounding the Vela pulsar.
However the intensity of this X-ray emission is on much lower level (see Figs. 6 and 7 
for comparison), which can be only partially explained by diffrent distances to these 
sources. The TeV $\gamma$-ray emission has been also detected from the region of $< 3$ pc 
around PSR 1706-44 (Kifune et al. 1995; Chadwick et al. 1998), with the intensity similar 
to that observed from the Vela nebula. Therefore, the spectral features of these two 
objects differ significantly, since if we take into account the difference in distances 
to these objects, $\sim 300$ pc to the Vela pulsar, and 1.8 kpc to PSR 1706-44
(Taylor \& Cordes~1993), PSR 1706-44 has much weaker X-ray emission and much stronger 
TeV $\gamma$-ray emission.  

As in the previous modelling, we assume that the initial period of this pulsar is also 
15 ms and the expansion velocity of the nebula 2000 km s$^{-1}$.
For the above parameters of the pulsar and nebula, the consistency with the observed 
period of the pulsar and dimension of the nebula is reached for the real age of this 
object equal to 16.000 yrs. In fact, due to the lower surface magnetic field of 
PSR1706-44, equal to $3.1\times 10^{12}$ G, and longer present period, 102 ms, 
the real age of the pulsar has to be closer to its characteristic age 17.400 yrs, than
in the case of the Vela pulsar. 

Now we are ready to calculate the expected spectrum from this nebula. 
Lower level of the synchrotron emission from the supernova remnant of the pulsar 
PSR 1706-44 suggest that the acceleration 
efficiency of leptons has to be lower than in the case of the Vela pulsar. 
The consistency with the X-ray emission from the nebula, reported by ASCA, is obtained 
for the efficiency of lepton acceleration $\xi = 0.02$, with the other model parameters 
fixed as in the case of the Vela and the Crab Nebulae (see thick dashed curve in Fig.~8).
However, the $\gamma$-ray emission from ICS of the diffusive low energy 
radiation (mainly MBR) by leptons is by a factor of a few below the
reported TeV $\gamma$-ray flux (see thick full curve in Fig.~8).
The $\gamma$-rays produced in the interactions of nuclei with the matter of the nebula 
can not explain the level of TeV $\gamma$-ray emission as well 
(dot-dashed curve in Fig.~8). 

The IC TeV emission could be enhanced if the additional source of soft radiation 
or high density matter is present in/or close to the nebula.
As already discussed in section 6.1, additional thermal 
emission with characteristic temperature $\sim 100$ K is observed from the Crab Nebula.
In the case of nebula around PSR 1706-44, the energy density of this infrared radiation
should be about a factor $\sim 5$ times higher than the MBR, in order to produce 
observable flux in the TeV $\gamma$-rays. We show the results of such calculations 
assuming that the temperature of this thermal radiation is 20 K (see thin dashed and 
full curves in Fig.~8). The required acceleration efficiency of leptons in this case is 
$\xi = 0.08$. However the presence of such strong infrared field inside the nebula 
is unclear. Finley et al.~(1998) suggest that likely candidates for such seed photons 
are IR background photons in the Galactic plane. 

\subsection{The Nebula MSH15-52 around PSR 1509-58}

The supernova remnant  MSH15-52, associated with the pulsar PSR 1509-58, has complex 
structure observed by many experiments in the radio and X-ray bands. 
The multiwavelenght spectrum of this nebula (see Fig.~9), extends from the radio (data
from du Plessis et al.~1995) through the X-rays (ROSAT - Trussoni et 
al.~1996, Beppo SAX - Mineo et al.~2001, RTXE - Marsden et al.~1997) up to the TeV 
$\gamma$-rays (Sako et al.~2000).
The pulsar and its nebula are sometimes identified with the historical supernova SN 185 
(Thorsett 1992). The distance to this object is put in the range 4.2 kpc 
(kinematics of H I, Caswell et al. 1975) up to 5.9 kpc (dispersion measure, Taylor \& 
Cordes 1993). We apply the value of 5.2 kpc derived by Gaensler et al.~(1999).
The present period of the pulsar is $\sim 150$ ms, its surface magnetic field 
is $1.5\times 10^{13}$ G, and
the characteristic age is $\sim 1700$ yr. These parameters are consistent with the 
initial pulsar period 15 ms, if the pulsar real age is close to its characteristic age. 
However, in order to fit the observed dimensions of this nebula, $5'\times 10'$
(Seward et al.~1984), we have 
to assume that the expansion velocity of the bulk matter in this nebula is much higher 
than applied in the modelling of other nebulae, and equal to 5000 km s$^{-1}$.
The initial mass of the nebula MSH15-52 is taken as in other cases equal to 
4 $M_{\odot}$. 

\begin{figure}
  \vspace{5.6truecm}
\includegraphics{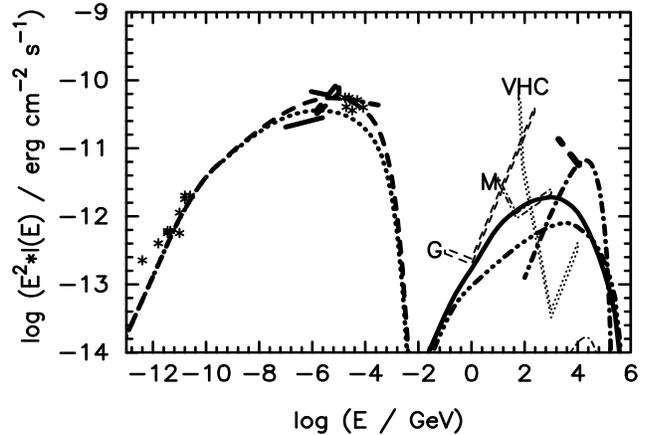}
  \caption{The multiwavelength spectrum of the nebula MSH15-52,  
from the radio up to the TeV $\gamma$-rays (see text for details). It is compared 
with the synchrotron and IC spectra produced by leptons which scatter the MBR and and 
synchrotron photons (dashed and dot-dot-dot-dashed curves, respectively). 
The synchrotron and IC spectra, calculated with the presence of 
additional target of infrared photons inside the nebula, are 
shown by the dotted and full curves, respectively. The initial parameters of the pulsar 
are: the period 
15 ms, the surface magnetic field $1.5\times 10^{13}$ G, and the nebula are: the mass 
$4 M_\odot$, and initial expansion velocity $5000$ km s$^{-1}$. The real 
age of the object is taken to be 1700 yrs. 
The $\gamma$-ray spectra from decay of $\pi^o$, produced by nuclei, are shown 
in the case of a supernova in the typical galactic medium with density of 0.3 cm$^{-3}$ 
(thin dot-dashed curve), and in the high density medium
with the average density of 300 cm$^{-3}$ (thick dot-dashed curve).
The 1 year sensitivity of the planned $\gamma$-ray satellite experiment is marked 
by the double dashed curve (G - GLAST ), and the
$5\sigma$ sensitivities (during 50 hours, $> 10$ photons) of the 
ground Cherenkov experiments are marked by the double dot-dashed curve (M - MAGIC), 
and double dotted curve (VHC - VERITAS, HESS, and CANGAROO III).}  
\label{fig9}
\end{figure}

For the above initial parameters of the pulsar PSR 1509-58 and its nebula, we have 
calculated the expected synchrotron and IC spectrum.
The consistency with the observed spectrum below 1 MeV is obtained for the differential 
spectral index of leptons equal to 2.1, the efficiency of acceleration of iron nuclei
$\chi = 0.85$, and the conversion efficiency of energy from nuclei to leptons equal to 
$\xi = 0.4$ (see thick dashed curve in Fig.~9). However, the IC spectrum is about an order 
of magnitude below the TeV $\gamma$-ray flux (Sako et al.~2000, see thick dashed line in
Fig.~9). 
This low $\gamma$-ray flux is caused by large adiabatic energy losses due to the fast 
expansion of the nebula. Even if we assume the existence of additional target of thermal 
infrared photons in the nebula (see e.g. Arendt~1989), with the temperature 100 K 
(as in the case of the Crab), and the energy density five times higher than the energy 
density of MBR, the IC spectrum can not explain the CANGAROO observations
(see thin full curve in Fig.~9).
The $\gamma$-ray emission from decay of $\pi^o$, produced in the interactions of nuclei 
accumulated inside this nebula (thick 
dot-dashed curve in Fig.~9), is much below the IC spectrum produced by leptons.

However there are evidences of a dense matter close to this supernova remnant.
The thermal, optical nebula, RCW 89, containing $H_\alpha$ line-emitting 
filaments with density $\sim 5\times 10^3$ cm$^{-3}$, coincides with the NW component of 
the remnant MSH15-52 (Seward et al.~1983). If the nuclei injected by the 
pulsar are captured by this high density filaments, then
the reported level of TeV $\gamma$-ray emission from this object might be due to
the $\gamma$-rays produced in the interactions of nuclei with the matter with the average
density of $\sim 300$ cm$^{-3}$ (see thin dot-dashed curve in Fig.~9). 
However, even in the most pessimistic case, the IC spectra produced by leptons in  
MSH15-52
should be detected by the new generation of Cherenkov telescopes at $\sim 1$ TeV
(see the sensitivity limit of the CANGAROO III and HESS telescopes marked by the double 
dotted curves in Fig.~9).

\subsection{The Nebula 3C58 around PSR J0205+6449}

The main features of the supernova remnant, 3C58, with its pulsar PSR J0205+6449, 
are quite similar to the Crab pulsar and Nebula. 
In fact, the PSR J0205+6449 period of 
65 ms, and estimated surface magnetic field of $3.6\times 10^{12}$ G, the distance  
of 3.2 kpc (Roberts et al.~1993), and the size of the nebula (a factor of two larger 
than the Crab Nebula) makes this object the closest to the Crab pulsar and Nebula.
Therefore, 3C58 should be considered as a good candidate for the TeV $\gamma$-ray source.
Moreover, the pulsar PSR J0205+6449 and its nebula 3C58 are probably very young, 
sometimes identified with the historic supernova in 1181 yr (Thorsett et al.~1992). 
The multiwavelength spectrum of this nebula extends from the  radio (Green~1986) up to 
the  X-rays (ASCA - Torii et al.~2000). The spectrum is additionally constrained by the 
upper limits in the infrared (Green \& Scheuer~1992), and the $\gamma$-rays by
the EGRET (Fichtel et al.~1994) and the Whipple (Hall et al.~2001). It is on a much 
lower level than observed from the Crab Nebula (see Figs. 6 and 10 for comparison).

Let us model the low energy spectrum of this nebula assuming that its real age is 
$\sim 820$ yrs, i.e. consistent with the age of the SN 1181 yr.
Then, the required initial period of the pulsar has to be $\sim 60$ ms and the expansion 
velocity of the nebula $\sim 5000$ km s$^{-1}$, in order to fit the observed parameters
of the pulsar and the dimension of the nebula. We calculate the synchrotron spectrum
from this nebula applying other parameters of the model consistent with that ones
derived for the Crab Nebula, i.e. the lepton spectrum,  $\chi = 0.85$, and $\xi = 0.5$. 
However, since the shape of calculated spectrum (see dotted curve in Fig.~10) is 
completely different than the observed one, it is rather unlikely that this PWNa can be
described properly by the above set of parameters.  

Therefore, we followed the proposition by Bietenholz et~al.~(2001), who argue that 
the real age of 3C58 is $\sim 5000$ yrs, consistent with the characteristic age 
of the pulsar PSR J0205+6449 (Murray et al.~2002). For this age, the  
initial period of the pulsar has to be 15 ms, and the expansion velocity 
1000 km s$^{-1}$, as proposed by Fesen~(1983). 
We have calculated the synchrotron spectrum from
3C58, applying the above initial parameters. The obtained fit to the 
radio - X-ray spectrum, assuming $\chi = 0.85$ and $\xi = 0.07$, is reasonable
(see dashed curve in Fig.~10). Note that, much lower value
for the efficiency of lepton acceleration, in respect to the Crab Nebula, is required in 
order to be consistent with the low X-ray flux from this nebula.
The calculated $\gamma$-ray emission from ICS (full curve) and from 
hadronic interactions (dot-dashed curve) is about $\sim 10^{-13}$ erg cm$^{-2}$ s$^{-1}$.
This is significantly below the $5\sigma$ sensitivity limit of the MAGIC telescope during
the 50 hour observation (double dot-dashed line in Fig.~10). However it is close to the 
$5\sigma$ sensitivity limit of the VERITAS telescopes during 50 hour observation 
(see double dotted broken line in Fig.~10).
Therefore, we conclude that predicted $\gamma$-ray emission above $> 100$ GeV from the 
nebula 3C 58, can be detected by the VERITAS telescopes if this source 
is observed longer than 50 hours.

\begin{figure}
  \vspace{5.6truecm}
\includegraphics{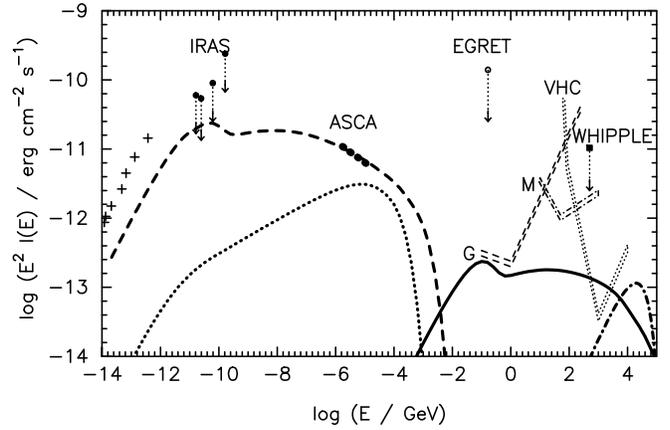}
  \caption{The multiwavelength spectrum of the nebula 3C58 (see text for details) is 
compared with the synchrotron spectrum (dashed curve) and the IC spectrum (full curve)
produced by leptons inside the nebula.
The initial parameters of the pulsar and the supernova are as in the case of the Crab 
Nebula, accept the surface magnetic field $3.6\times 10^{12}$ G, and the 
age of 5000 yrs. The $\gamma$-ray spectrum from decay of pions is shown by the 
dot-dashed curves. 
The dotted curve shows the synchrotron spectrum from the nebula assuming that this
object is connected with the supernova in 1181 yr, which require the initial
period of the pulsar 60 ms, and the expansion velocity of the nebula 5000 km s$^{-1}$.
The sensitivities of the planned experiments are shown as in Fig.~9.}   
\label{fig10}
\end{figure}
\subsection{The Nebula CTB 80 around PSR 1951+32}

The pulsar PSR 1951+32 has a short period, 39.5 ms, large characteristic age of 
$1.1\times 10^5$ yr, and a relatively weak surface magnetic field $4.9\times 10^{11}$ G.
It is inside the supernova remnant CTB 80, which consists of a $10'\times 6'$ compact 
radio nebula immersed in a radio shell-like extended source with a diameter of $30'$
(e.g. Velusamy, Kundu, Becker~1976, Angerhofer et al.~1981). The differential radio 
spectrum has the average spectral index of -0.6. The estimated dynamic age of the 
shell-like nebula $9.6\times 10^4$ yr (for the distance 2.5 kpc) agrees well with the 
characteristic age of the pulsar (Koo et al.~1990). 
In the X-rays, the nebula is composed of a compact nebula of $\sim 1'$ and the diffuse
nebula with the radius of $\sim 5'$ (Safi-Harb et al. 1995). The differential X-ray
spectrum of the compact nebula is equal to 2.1 and luminosity $3.9\times 10^{33}$ ergs 
s$^{-1}$. When modelling this nebula, we consider two possibilities. 

In the first one, we assume
that these objects are at the distance of 2.5 kpc and have the real age close to the 
pulsar's characteristic age. Then, the observed parameters of the pulsar and nebula are 
consistent with the initial period of the pulsar of 15 ms and the initial expansion 
velocity of the shell-like nebula equal to 2000 km s$^{-1}$, as in the case of the Crab 
Nebula. The synchrotron spectrum calculated for these parameters fits quite well
to the observations from the radio up to the X-rays (see thick dashed curve in Fig.~11). 
However, in order to obtain better fit we slightly modified the parameters derived for the
Crab Nebula, by changing the efficiency of positron acceleration to $\xi = 0.15$, and the 
spectral index of injected positrons to 2.3. The IC spectrum produced by leptons inside 
the compact nebula is shown by the thick full curve in Fig.~11. This $\gamma$-ray 
emission should be 
easily detected by the GLAST detector in the GeV energies and by the Cherenkov telescopes 
(MAGIC, VERITAS) at TeV energies. The $\gamma$-ray emission from decay of pions, 
produced in collisions of nuclei, is negligible for the nebula with the age $\sim 10^5$ 
yrs (see thick dot-dashed curve in Fig.~11).

\begin{figure}
  \vspace{5.6truecm}
\includegraphics{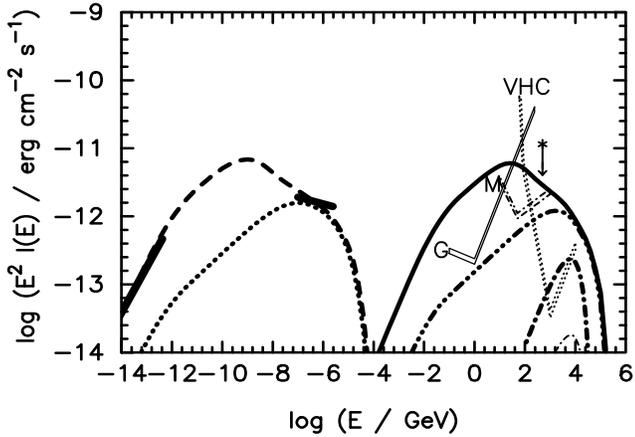}
  \caption{The multiwavelength spectrum of the nebula CTB 80, around PSR 1951+32,
(see text for details) is compared with the synchrotron spectrum (thick dashed curve) 
produced by leptons in the inner compact nebula, and the IC spectrum produced by 
leptons which comptonize the MBR in the compact nebula (full curve).
The initial parameters of the pulsar and the supernova are as in the case of the Crab 
Nebula, accept the acceleration efficiency of leptons, $\xi = 0.15$ and their spectral 
index equal to 2.3.
The surface magnetic field of the pulsar is equal to $4.9\times 10^{11}$ G, and the 
real age is $10^5$ yrs. The $\gamma$-ray spectrum from decay of pions produced by 
nuclei within the shell-like radio nebula is shown by the thin dot-dashed curve.
If the nebula is the remnant of the supernova at the year 1408, then the expected
synchrotron and IC spectra from leptons are shown by the dotted and 
dot-dot-dot-dashed 
curves and the spectrum from decay of pions by the thick dot-dashed curve.}   
\label{fig11}
\end{figure}

In the second modelling, we follow the claims that 
the nebula CTB 80 is the remnant of the recent supernova at the year 
1408, found in the Chinese recorders (Strom et al.~1980, Wang \& Seward~1984). 
Then, the initial period of the pulsar has to be very close to the observed one,
i.e. equal to $\sim 39.35$ ms, due to the low surface magnetic field. In this case, 
the radio shell-like extended nebula with a diameter of $30'$ can not be related to the 
CTB 80 at all. To fit the observed size of the compact nebula, the expansion velocity of 
the bulk matter of this supernova has to be $\sim 1000$ km s$^{-1}$.
We have tried to describe the observed X-ray spectrum of the compact nebula around 
PSR 1951+32, assuming that the mass of the nebula is also equal to 4 $M_\odot$,
$\chi = 0.85$, $\xi = 0.2$, and the index of the power law spectrum of injected leptons 
equal as before to 2.3. The corresponding synchrotron and IC spectra from leptons are
shown in Fig.~11 by the thin dashed and full curves. Although, we are able to fit
reasonably well the observed X-ray spectrum, the lower part of synchrotron spectrum 
is significantly below the level of the radio emission. However, since the radio 
emission corresponds to a much larger region than the X-ray compact nebula, 
connected rather with the extended nebula with the radius of a $10'\times 6'$, we 
can not reject the connection of the CTB 80 with the supernova at the year 1408.   
We have also calculated the $\gamma$-ray emission from decay of pions assuming the 
above parameters of the nebula. These emission can slightly contribute to the IC spectrum
at $\sim 10$ TeV.

\section{Conclusion}

We have applied the model for the acceleration of positrons in resonant interactions
with the heavy nuclei (see e.g. Arons 1998)
to construct a time dependent model for the radiation processes in the PWNe.
We calculate self-consistently the amount of relativistic nuclei and leptons inside the
PWNa as a function of time, measured from the supernova explosion, i.e. formation of a 
pulsar.
The model describes well the lower energy part of the broad band spectrum observed 
from the Crab Nebula by the synchrotron spectrum of leptons. The 
$\gamma$-ray spectrum below a few TeV is described by the inverse Compton spectrum from 
scattering of low energy synchrotron photons, microwave background radiation, and 
infrared photons by leptons, and above $\sim 10$ TeV by the $\gamma$-rays from decay
of pions produced in hadronic collisions of nuclei.
From the fitting to the Crab Nebula spectrum, we derive the free parameters of this model. 
These parameters are used as frequently as possible in modelling of other nebulae
from which the TeV $\gamma$-ray emission have been reported. We consider two southern sky 
nebulae, around the Vela pulsar (89 ms period) and around PSR 1706-44 (102 ms period), 
which have quite similar parameters. By using these same initial periods of the pulsars 
and the expansion velocities of their nebulae, as obtained from modelling of the Crab 
pulsar, we get the observed values
for these two pulsar's periods and dimensions of their nebulae. 
We have obtained reasonable fitting to the X-ray and TeV $\gamma$-ray flux from the
Vela nebula for the model of acceleration of leptons as defined in the case of 
the Crab Nebula. However, the X-ray luminosity of the nebula around PSR 1706-44 is  
significantly lower than that observed from the Crab Nebula. Therefore, we have to 
apply much lower efficiency of acceleration of leptons in this nebula in order to explain 
the X-ray spectrum by synchrotron process. In such a case the calculated 
TeV $\gamma$-ray
flux from IC scattering of the MBR by leptons is a few times below the
fluxes reported by the CANGAROO and the Durham groups (Kifune  et al. 1995, 
Chadwick et al. 1998). 
These observations can be explained by this model if we postulate the existence
of additional photon target for leptons inside the nebula, e.g. thermal infrared photons.   

Next, we interpret the southern sky nebula MSH15-52 (containing the $\gamma$-ray 
pulsar PSR 1509-58), which has been marginally detected at the TeV energies by the 
CANGAROO
telescope (Sako et al. 2000). The observed period of PSR 1509-58 is consistent with
the initial period of 15 ms, as in the case of the Crab pulsar. But, the dimensions and 
the age of the nebula argue for a much higher expansion velocity of this nebula.
We obtained reasonable fit to the radio and X-ray emission from MSH15-52. However,  
calculated self-consistently TeV $\gamma$-ray flux is about an order of magnitude 
below that one reported by Sako et al. (2000). This nebula is close to the dense
optical nebula RCW 89. If significant amount of nuclei injected by the pulsar
is captured by these high density region, then the $\gamma$-rays from decay of pions,
produced in hadronic collisions, can be explain the level of reported TeV emission.  
Therefore, we suggest that the TeV emission reported by the CANGAROO can have hadronic
origin in contrary to the calculations of Du Plessis et al. (1995), who argue for the 
dominance of leptonic emission also in this energy range.

Based on the experience reached from modelling of the objects for which TeV $\gamma$-ray 
emission has been reported, we try to predict the level of the high energy $\gamma$-ray 
emission from other nebulae, which may become good potential targets 
for the next generation of the GeV-TeV $\gamma$-ray telescopes. 
We concentrate on the two northern sky nebulae: 3C58 and CTB 80, which contain fast 
pulsars PSR J0205+6449 (65 ms) and PSR 1951+32 (39.5 ms), respectively.
It is found that 3C58 is too weak to be detected in the GeV energy range by the GLAST 
detector and lays on the sensitivity limit of the VERITAS telescope in the TeV energies.
From another site, CTB 80 should be easily detected by the GLAST detector and the MAGIC 
and VERITAS telescopes.

The experience reached from these modelling convince us that it is rather difficult to
describe consistently the high energy processes in the PWNe by a simple model with
a small number of parameters. It looks that individual sources require different 
approach which is determined by the specific conditions around and inside the nebula,
like for example the presence of a dense matter or additional targets of soft photons
appart from the synchrotron radiation produced inside the nebula and MBR.

Our modelling base on the assumption that most of the rotational energy of the pulsars
are injected in the form of relativistic nuclei. Heavy nuclei, which escape from the PWNe,
can contribute significantly to the cosmic ray spectrum at energies above the knee region
(Bednarek \& Protheroe 2002), at which the enhancement of heavy composition in the 
cosmic ray spectrum is reported by some experiments. 
Another independent test on the importance of injection of nuclei by the pulsars and on 
the validity of the considered here model for the acceleration of leptons can be 
provided by the observations of the high energy neutrinos from the PWNe by the future 
neutrino telescopes (IceCube, ANTARES, NESTOR). We have considered this problem in an 
accompanying paper (Bednarek 2003). It is shown there that, in contrary to previous 
estimates (Guetta \& Amato 2002), only very young nebulae (of the Crab type) can produce 
observable neutrino signals in a 1 km$^2$ detector. 

\begin{acknowledgements}
This work is supported by the Polish KBN grants No. 5P03D 025 21 and 
PBZ-KBN-054/P03/2001. 
\end{acknowledgements}

\end{document}